\documentclass[aps,prl,twocolumn,superscriptaddress,nofootinbib,]{revtex4-1}
\usepackage{graphicx}
\usepackage{float}
\usepackage{longtable}
\usepackage{array}  
\usepackage{color}
\usepackage{ulem}

\begin{document}
\title{Measurement of  $\gamma$ rays from the giant resonances excited by $\rm^{12}C$(\textit{p,p}$^\prime$) reaction at 392 MeV and 0$ ^{\circ}$ }

\author{M. S. Reen}
\email[e-mail adress: ]{mandeepsingh@okayama-u.ac.jp}

\author{I. Ou}
\author{T. Sudo}
\author{D. Fukuda}
\author{T. Mori}
\author{A.Ali}
\author{Y. Koshio}
\author{M. Sakuda}

\thanks{M.Sakuda}
\email{sakuda-m@okayama-u.ac.jp}
\affiliation{Department of Physics, Okayama University, 700-8530 Okayama, Japan}

\author{A. Tamii}
\thanks{A. Tamii}
\email{tamii@rcnp.osaka-u.ac.jp}
\author{N. Aoi}
\author{M. Yosoi}
\author{E. Ideguchi}
\author{T. Suzuki}
\author{T. Yamamoto}
\affiliation{Research Center for Nuclear Physics (RCNP), Osaka University, 567-0047 Osaka, Japan}

\author{C. Iwamoto}
\affiliation{Center for Nuclear Study, University Of Tokyo (CNS) RIKEN campus, 351-0198 Saitama, Japan}

\author{T. Kawabata}
\affiliation{Department of Physics, Osaka University, 567-0043 Osaka, Japan.}

\author{S. Adachi}
\author{M. Tsumura}
\author{M. Murata}
\author{T. Furuno}
\affiliation{Department of Physics, Kyoto University, 606-8502 Kyoto, Japan}

\author{H. Akimune}
\affiliation{Department of Physics, Konan University, 658-8501 Hyogo, Japan}

\author{T. Yano}
\affiliation{Department of Physics, Kobe University, 657-8501 Hyogo, Japan}

\author{T. Suzuki}
\affiliation{Department of Physics, Nihon University, 156-8550 Tokyo, Japan}

\author{R. Dhir}
\affiliation{Department of Physics and Nanotechnology, SRM University, 603203 Kancheepuram, India}


\date{\today}

\begin{abstract}

We measured both the differential cross section ($\sigma_{p,p^\prime}$ $=d^2\sigma/d\Omega dE_{x}$) and the $\gamma$-ray emission probability
($R_\gamma(E_x)$ $=\sigma_{p,p^\prime\gamma}$/$\sigma_{p,p^\prime}$) from the giant resonances excited by  $\rm^{12}C$(\textit{p,p}$^\prime$) reaction at 392 MeV and 0$^\circ$, using a magnetic spectrometer  and an array of NaI(Tl)  counters. The absolute value of $R_\gamma(E_x)$ was calibrated by using the well-known $\gamma$-ray emission probability from $\rm^{12}C^*  ( 15.11$ MeV, $ 1^+$, $T=1$) and $\rm^{16}O^*  ( 6.9$ MeV, $2^+$, $T=0$) states  within 5\% uncertainty. We found  that $R_\gamma(E_x)$ starts from zero at $E_x=16$ MeV, increases to a maximum of  53.3$\pm$0.4$\pm$3.9\% at $E_x=27$ MeV and then decreases. We also compared the measured values of $R_\gamma(E_x)$ with statistical model calculation based on the Hauser-Feshbach formalism in the energy region $E_x=$ 16-32 MeV and discussed the features of $\gamma$-ray emission probability quantitatively.

 \end{abstract}


\maketitle

\section{I. Introduction}

Carbon  is the fourth most abundant element by mass in the solar system \cite{abundance} after hydrogen, helium, and oxygen, and $\rm ^{12}C$ is its most abundant (98.9\%) isotope.  Thus, it has been used as a target material in the form of  organic liquid scintillators in many large-scale neutrino experiments designed to detect low-energy neutrinos ($E_\nu$\textless 100 MeV) \cite{Reines, KARMEN,KARMEN2, LSND, KamLAND}. These detectors must be massive to compensate the extremely small neutrino cross section ($\approx 10^{-42}$ $\rm cm^2$). One of the most interesting applications 
is the detection of neutrinos from supernova explosion in our Galaxy \cite{Bethe, Koshiba}. The main reaction for neutrino detection is the charged-current  (CC) anti-neutrino reaction with a proton  ($\bar{\nu}_e+ p\to e^+ + n$), also known as the inverse $\beta$-decay reaction (IBD). Of special interest is the neutral-current (NC) neutrino or anti-neutrino inelastic scattering  with $\rm ^{12}C$, followed by the emission of $ \gamma$ rays that can be observed with the detector \cite{Donnelly}. This process is of a special interest because the cross section is significant enough to be detected and is independent of neutrino oscillations. 

The first observation of $\rm ^{12}C( \nu,  \nu ')^{12}C^*( 15.11$ MeV, $1^+$, $T=1$) reaction with 15.11-MeV  $\gamma$ ray came from the KARMEN experiment  \cite{KARMEN,KARMEN2} with a neutrino beam. The observation was based on the detection of the electromagnetic decay of  $\rm ^{12}C$ excited by neutral current interactions. The $\gamma$-ray emission probability  ($  \Gamma_ {\gamma}/ \Gamma$) of excited states of $\rm ^{12}C$ below the proton separation energy  ($S_p=16.0$ MeV) has been well measured \cite{TableIsotope}. However, the giant resonances appear  above  the separation energy and they decay mainly hadronically via particle emission  ($p, n, d$ and $ \alpha$) to the daughter nuclei. Although they decay  mainly to the ground state of the daughter nuclei ($\rm ^{11}B$, $\rm ^{11}C, etc.$), some of these decays are to excited states. If these excited states are below the particle emission threshold in $\rm ^{11}B$  ($S_{p'}= 11.2$ MeV) or $\rm ^{11}C$  ($ S_{p'}= 8.7$ MeV), they decay by $\gamma$-ray emissions. Kolbe \textit{et al.} and Langanke \textit{et al.} \cite{Langanke,kolbe} proposed the above decay mechanism of giant resonances and estimated the NC neutrino and anti-neutrino reaction cross sections for $\rm ^{12}C$ and $\rm ^{16}O$. 

 They stressed  the importance of measuring NC events, since they are more sensitive to $\nu_{\mu}$ and $\nu_{\tau}$ neutrinos than to $\nu_e$ neutrinos\footnote{This statement is based on the past predictions for the average neutrino enegies~\cite{Bethe,Qian}. The more recent calculations on neutrino spectra from supernova explosion suggest that the average neutrino energies are not very different between neutrino flavours~\cite{Buras}.}. However, there are no experimental measurements of  $\gamma$ rays from the giant resonances of  $\rm ^{12}C$.

In this paper, we report the first measurement of $\gamma$ rays from the excited states of  $\rm^{12}C$, including giant resonances in the energy region $ E_x=$ 16-32 MeV.

\section{II. Experiment}
\begin{figure*}[t!]
\centering
\includegraphics[width=20cm,bb=0 0 2100 806]{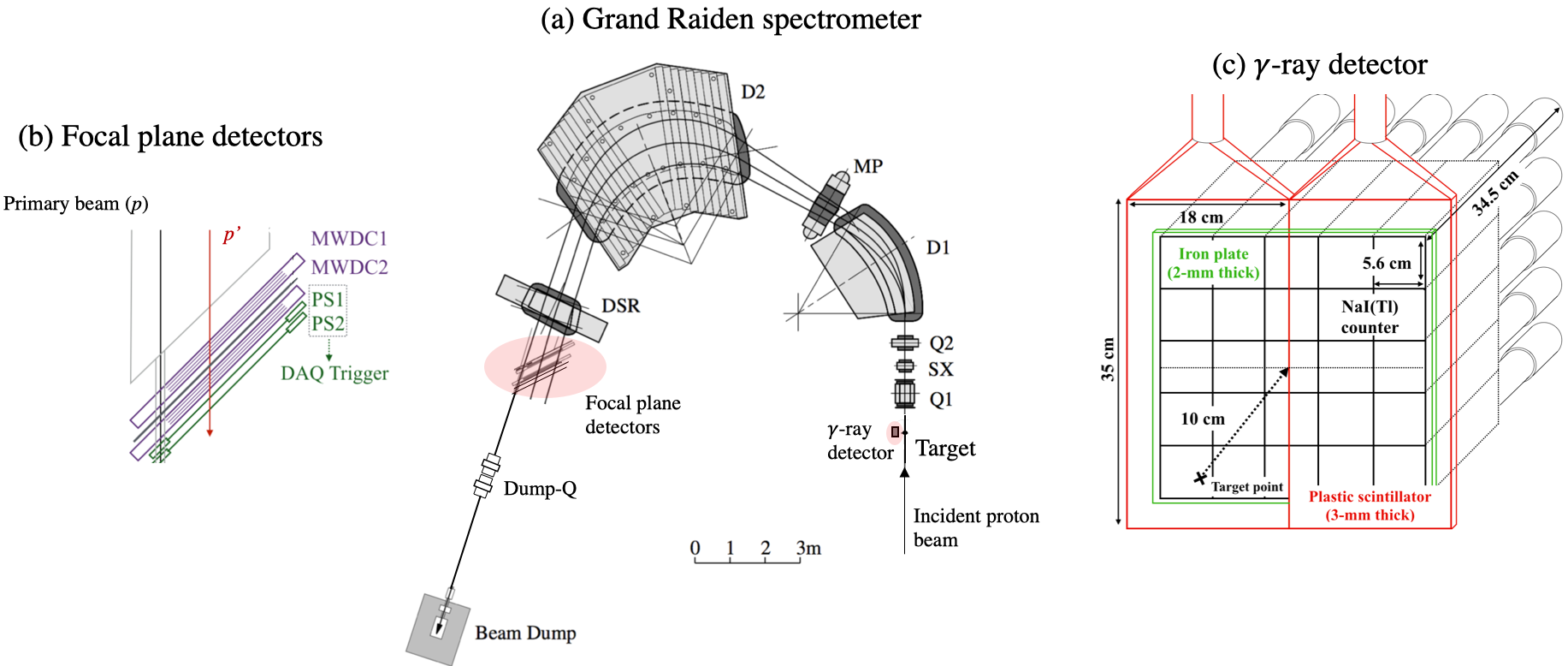}
\caption{(a) Grand Raiden spectrometer in the experimental setup at 0$^\circ$, (b) Focal plane detectors, and (c) $\gamma$-ray detector.  \label{fig:GR}}
\end{figure*}

The experiment (E398) to measure the $\gamma$ rays emitted from giant resonances in $\rm ^{12}C$  was carried out at the Research Center for Nuclear Physics  (RCNP), Osaka University. 
An unpolarized proton beam at 392 MeV bombarded a natural carbon ($\rm ^{nat}C$) target with a beam bunch interval of 59 ns. The scattered protons were measured around 0$\rm^\circ$ and were analyzed by the high-resolution magnetic spectrometer Grand Raiden (GR) \cite{peter}. The layout of (a) Grand Raiden (GR) spectrometer, (b) Focal plane detectors, and (c) $\gamma$-ray detector is shown in Fig. \ref{fig:GR}.

\subsection{A. Grand Raiden magnetic spectrometer}
Two  multi-wire drift chambers (MWDC)  were placed at the focal plane of the GR system followed by two plastic scintillators (PS 1 and 2).  Each of PS1 and PS2 was coupled with two photo-multiplier tubes (PMT) from each side. A fast trigger (PS trigger) was generated by the coincidence of the discriminator signals of PS1 and PS2 for the data-acquisition (DAQ) system. Signals from the MWDCs were  pre-amplified and discriminated by a LeCroy 2735DC board and  the timing information of the wires was digitized by LeCroy 3377 time-to-digital converter (TDC). The details of the DAQ system were described elsewhere~\cite{tamii3} and only the components necessary for the present paper are described here. The MWDCs measure a charged-particle track at the focal plane of the GR spectrometer and  were used to measure the excitation energy of the target nucleus ($E_x = E_p-E_{p'}$) and the scattering angle of protons ($\theta _p$) at the target position. The spectrometer covered the scattering angle range of 0$\rm ^\circ$\textless$\theta_p$\textless 3.5$\rm ^\circ$. The beam current was monitored by a Faraday cup located at the beam dump and the typical beam intensity was 0.5-1.5 nA. An energy resolution of 120 keV (FWHM) was achieved at  $ E_x=$ 15.1 MeV. Details of the GR spectrometer have been described elsewhere~\cite{GrandRaiden,tamii2}.

\subsection{B. $\gamma$-ray detector}
A $\gamma$-ray detector was made from an array of 5$\times$5 NaI(Tl) counters. 
One NaI(Tl) counter was made up of a 5.1 cm$\times$5.1 cm$\times$15.2 cm crystal and  a photo-multiplier (Hamamatsu  R980)
whose photo cathode (3.8 cm in diameter) was attached to one end of the crystal. The crystal  was contained in an air-tight 1mm-thick aluminum case and a thin white reflective sheet was inserted between the crystal and aluminum case. Thus, one NaI(Tl) counter has a total size of 5.6 cm$\times$5.6 cm$\times$34.5cm.  Each photo-multiplier was covered by a $\mu$ metal. The  $\gamma$-ray detector array was placed at $\theta =90\rm ^\circ$ with respect to the beam direction and at a distance of 10 cm from the target. 
The front face and sides of the detector  were covered by a 2-mm thick iron plate to suppress low-energy beam-induced and ambient $\gamma$ rays less than 200 keV. 
Two 3-mm thick plastic scintillators (veto counters) were attached in front of the iron plate and the NaI(Tl) counters to separate the background caused by charged particles directly entering the $\gamma$-ray detector.  The scintillation light was measured from one end of the scintillator by a photo-multiplier (Hamamatsu H6410) through an acrylic light guide.

 For each PS trigger, both the ADC (charge information) and TDC (time information) of each PS counters were recorded. The GAM signal is defined as the sum of discriminator signals of all NaI(Tl) counters. A GAM trigger was generated by taking the coincidence of the PS trigger and the GAM signal, and was used for the data acquisition of ADC and TDC of NaI(Tl) counters and veto counters. Those signals were digitized and recorded by LeCroy FERA and FERET systems.

\begin{figure}[ht!]
\centering
\includegraphics[width=14cm,bb=0 0 1520 700]{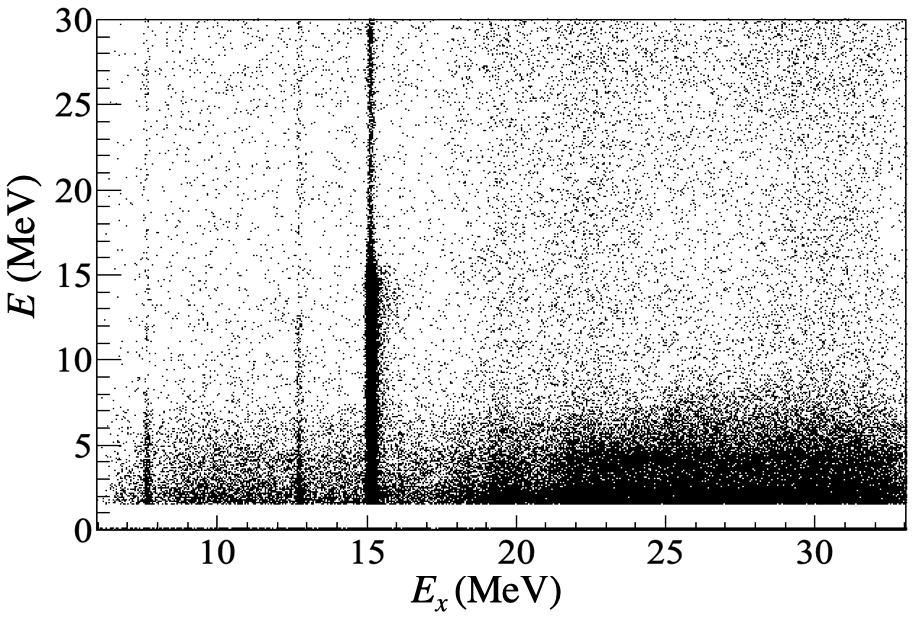}
\caption{Two-dimensional histogram with E ($\gamma$-ray energy measured by the detector) at y axis and $E_x$ at x axis with each point showing a coincidence event. Accidental background has not been subtracted.\label{fig:coin}}
\end{figure}

\begin{figure}[hb]
\centering
\includegraphics[width=9.5cm,bb=0 0 1420 700]{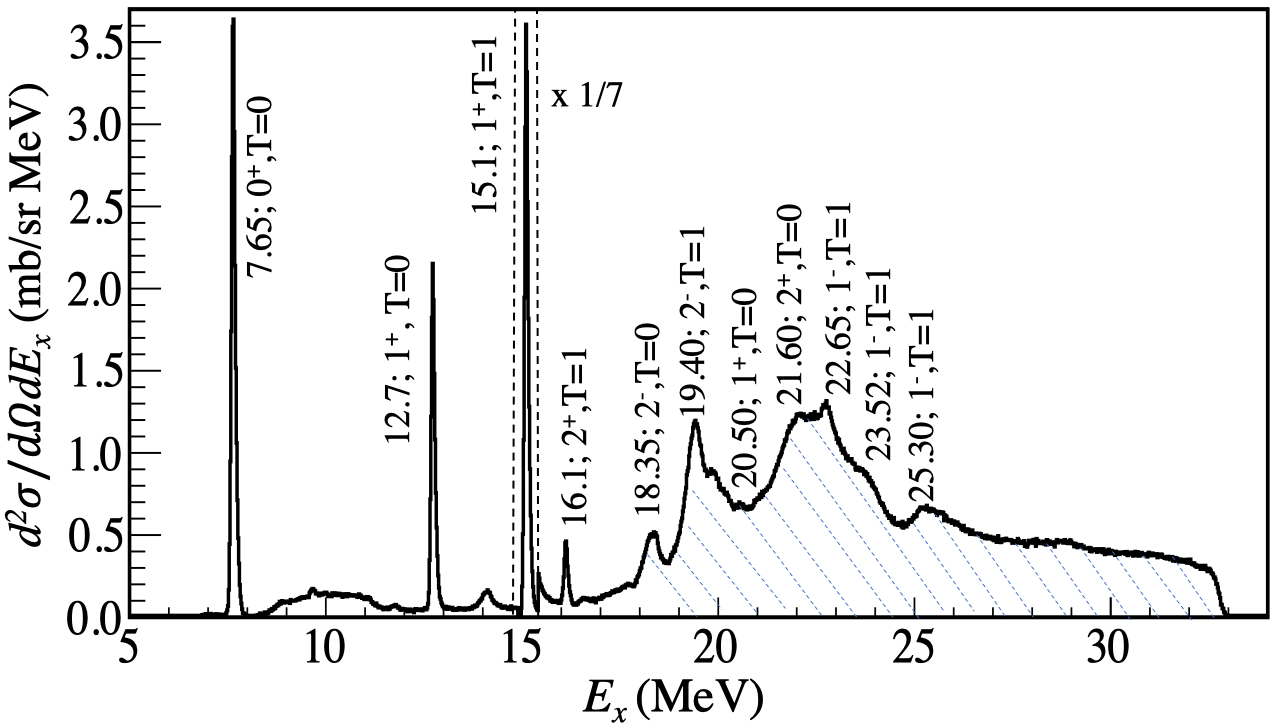}
\caption{Double differential cross section of the $\rm^{12}C$(\textit{p,p}$^\prime$)  reaction at $  E_p=392$ MeV and $ \theta =  0^{\circ}$. The bin width is 0.02 MeV.\label{fig:cross}}
\end{figure}

\begin{figure*}[ht]
\centering
\includegraphics[width=31cm,bb=0 0 2100 406]{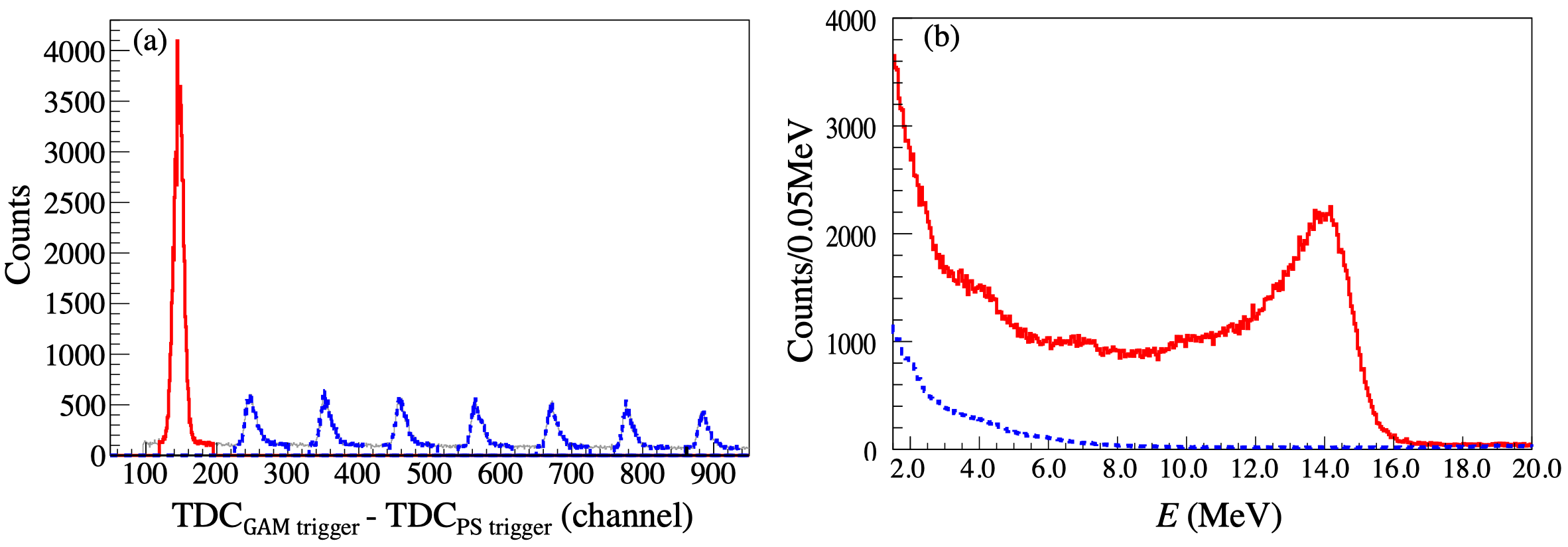}
\caption{(a) Time difference between GAM trigger and PS trigger. (b) $\gamma$-ray energy spectrum (red solid line) and background  spectrum (blue dotted line) for the 15.11-MeV state ($1^+$, $T=$1) of $\rm ^{12}C$. \label{fig:gammabg}}
\end{figure*}

While the initial energy calibration for all NaI(Tl) counters was performed by using a $\rm ^{60}Co$ source before the experiment, the energy response of the NaI(Tl) counters decreased gradually under the exposure of beam due to irradiation by beam-induced particles. Therefore, we calibrated the energy response of each NaI(Tl) counter for each run (typically 2 hours) by using the following in-situ $\gamma$ rays, $\rm ^{12}C$(15.11 MeV, $ 1^+$), $\rm ^{11}B$(2.12 MeV, $1/2^-$) and 1.37 MeV from $\rm{^{24}Mg^*}$. The 1.37 MeV $\gamma$ ray was induced by secondary interactions with the aluminum of the chamber surrounding the target. The mean energy of 1.37 MeV was determined by the nearby Germanium counter. During the in-situ calibration, we found that 15 downstream counters had poor energy resolution, so we used only the other 10 upstream counters. The energy resolution $ \sigma (E)/E$ of each 10 upstream counters was 5\% at 2 MeV and 3\% at 15 MeV. The experiment was conducted with three beam intensities, 0.5, 1.0 and 1.5 nA but the gain variation was least for the 0.5 nA dataset. Therefore, that dataset was used for the $\gamma$-ray analysis.

\subsection{C. Scattered proton and $\gamma$-ray coincidence measurement}

The main feature of this experiment is to measure both the excitation energy $E_x$ by the GR spectrometer and the $\gamma$-ray energy ($E$) by  the NaI(Tl) counters. 
We define the $\gamma$-ray energy ($E$) as the sum of the pulse height measured in the upstream 10 NaI(Tl) counters. Thus, we study both the cross section ($\sigma_{p,p^\prime}$ $=d^2\sigma /d\Omega d E_x$) and the $\gamma$-ray emission probability
($R_\gamma(E_x)=\sigma_{p,p^\prime\gamma} / \sigma_{p,p^\prime}$) from the giant resonances.   Figure \ref{fig:coin} presents the spectra of the excitation energy ($E_x$) and the measured $\gamma$-ray energy ($E$)  for the coincidence events between the PS trigger and the GAM trigger.   

By taking a typical  $ \gamma $ ray 15.11 MeV, we explain  in the following how we measured the $\gamma$-ray energy ($E$) and estimated the accidental background by using both ADC and TDC informations for each $E_x$ interval. The time difference between the GAM trigger and the PS trigger is plotted in Fig.~\ref{fig:gammabg} for $ \gamma $ rays from $\rm^{12}C($15.11 MeV, $ 1^+, T=1$). Events in the prominent first peak (red) were selected as coincidence events between the two triggers, whereas those in the other peaks were selected as accidental background. Pulse intervals of 59 ns correspond to the bunch structure of the beam. Thus, we obtained the energy deposit $E$ for the signal (red line) and the background  (blue line) for $E_x $=15.11 MeV  in Fig.~\ref{fig:gammabg}(b). The details of the analysis will follow in Sections III and IV.

\section{III. Analysis of scattered protons}
\subsection{A. ($p,p^\prime$) differential cross section}


\begin{table}[hb]
\begin{tabular}{>{\centering \arraybackslash}p{3.5cm}>{\centering \arraybackslash}p{1.7cm}} 
\hline
 \hline
Variable  & Value \\
\hline
Tracking efficiency  ($\eta$) & 1\%\\
Solid angle  ($ \Omega $) & 3\% \\
Beam charge  (Q) & 3\%\\
Target thickness  (t) &2\%\\
Background subtraction& 3\%\\
\hline
Total&6\%\\

\hline
\hline
\end{tabular}
\caption{Systematic uncertainties in the measurement of differential cross section.}\label{tab:sys1}
\end{table}

The double differential cross section is given as

\begin{eqnarray}
 \sigma _{p,p'} \equiv \frac{d^2\sigma}{d\Omega dE_x} = J\frac{N_{E_x}}{\Delta E_x }\frac{1}{\Omega}\frac{1}{L\eta}\frac{e}{Q}\frac{A}{N_A \rho },
\label{eqn:emitpd}
\end{eqnarray}
where  $J$ is the Jacobian for the transformation from laboratory frame to c.m. (center of mass) frame (0.81), $\eta$ is the tracking and trigger efficiency  (0.91), $L$ is the DAQ live time, e is the elementary charge  (C), $Q$ is the total beam charge  (C), and $ N_{E_x}$ are the number of excitation events in the energy range $ { E_x}$ and $\ {E_x + \Delta E_x}$ obtained after subtracting the background. The  detailed procedure for background subtraction was provided in Ref.~\citep{tamii2}. Furthermore, $A$ is the atomic weight  (g/mol), $ N_A$ is Avogadro constant, 
and $ \rho $ is the areal density (36.3 mg/$\rm  cm^2 $). The spectrometer acceptance  was not symmetrical with respect to the horizontal and vertical directions ($ -9$ mrad $\leq\theta_x \leq 0 $ mrad, $|\theta_y| \leq 43 $ mrad). The events were chosen within a solid angle ($\Omega$) of 0.77 msr.

\begin{figure}[h!]
\includegraphics[width=8.5cm,scale=0.2]{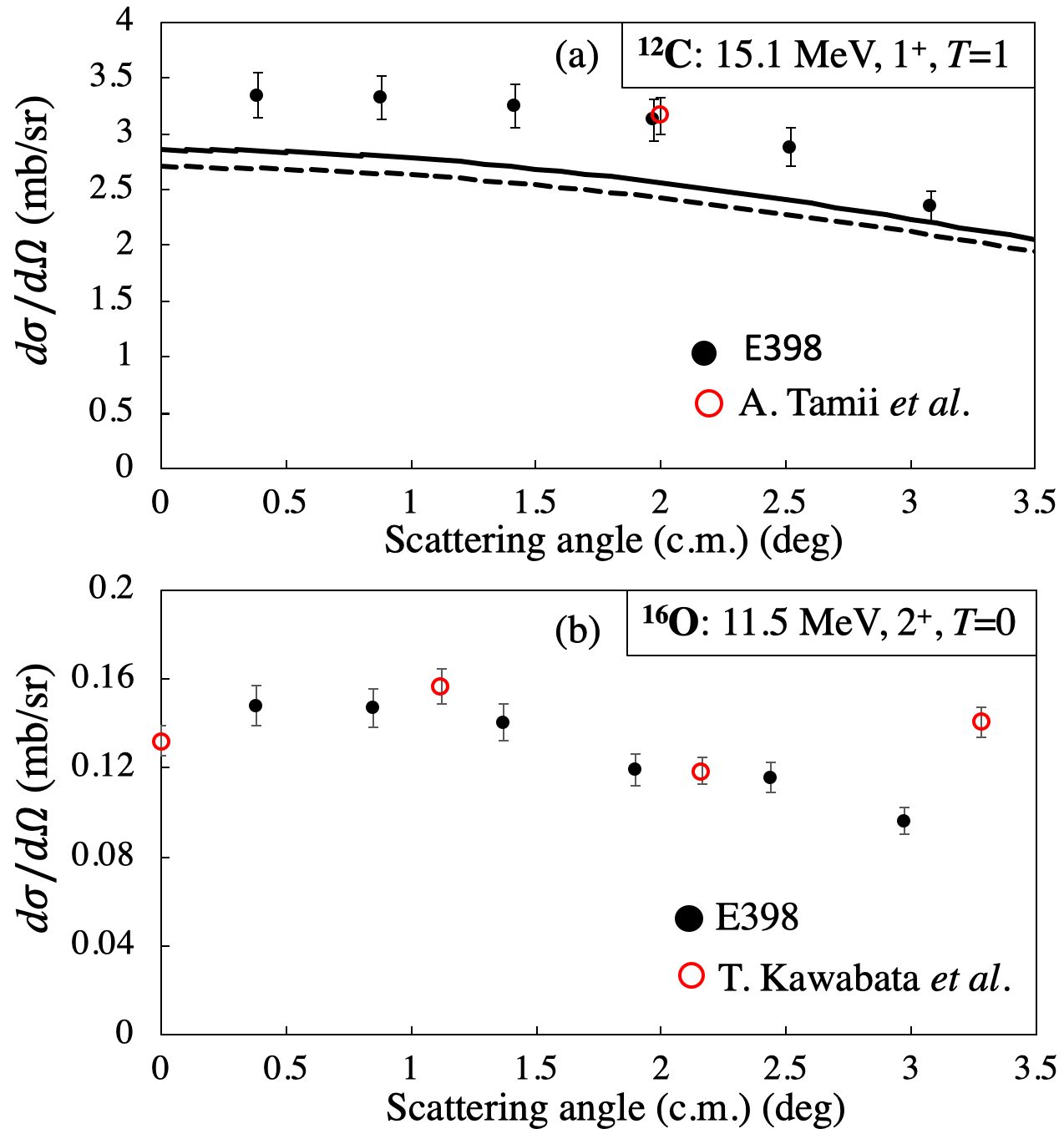}
\caption{(a) Differential cross section of the $\rm^{12}C$(\textit{p,p}$^\prime$) reaction as a function of scattering angle (black circles) and comparison with previous experiment~\cite{Tamii} (red open circles). Solid (SFO) and dashed (Cohen-Kurath) lines are the DWBA calculation results for the transitions to 15.1-MeV state (see text for details). (b) Differential cross section for $\rm^{16}O$(\textit{p,p}$^\prime$) reaction as a function of scattering angle and comparison with previous experiment~\cite{kawabata}. \label{fig:comp2}}
\end{figure}

The measured cross section of $\rm^{12}C$(\textit{p,p}$^\prime$)  is shown in Fig.~\ref{fig:cross}. Giant resonances are clearly seen in the spectrum. We list  the excitation energies $E_x$,  spin-parities ($J^{\pi}$), and isospin ($T$) of the known resonances in Table~\ref{tab:crossfit}. 
We show the differential cross section for $\rm^{12}C$(15.11 MeV, $1^+, T=1$) and $\rm ^{16}O$(11.5 MeV, $  2^+, T=0)$ in Fig.~\ref{fig:comp2}, demonstrating the consistency of our cross section with those of previous experiments performed with the same GR spectrometer  at the same beam energy~\cite{Tamii, kawabata}. Our cross section measurements of $\rm^{16}O$(11.5 MeV, $2^+, T=0)$ were performed during the same experiment with a cellulose ($\rm C_6H_{10}O_5$) target. Both of our measured cross sections are consistent with those measured in previous experiments within the systematic uncertainty of 6\%.

\begin{figure}[ht!]
\centering
\includegraphics[width=38cm,bb=0 0 2240 710]{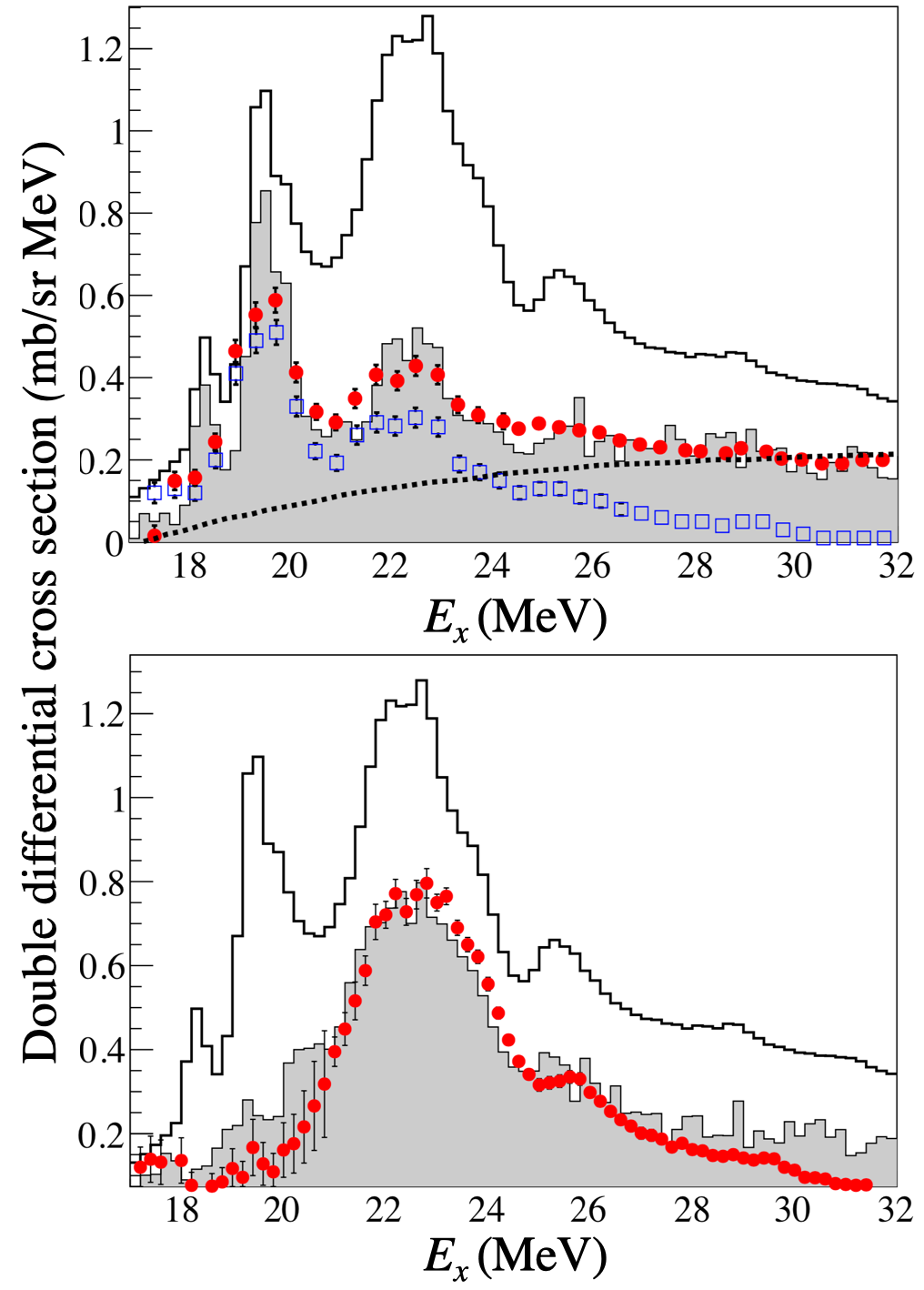}
\caption{(a) Spin-flip component $\Sigma \hspace{0.1cm} d^2\sigma/d\Omega dE_x$ (shaded region) is compared with  $d^2\sigma/d\Omega dE_x$ (solid line).  The spin-flip cross section for $\rm^{12}C$(\textit{p,n})$\rm ^{12}N$ reaction (blue open squares), the contribution of quasifree process (dotted line), and their sum (red circles) are obtained from Ref.~\cite{dozono}. (b) Non-spin-flip component $(1-\Sigma) \hspace{0.1cm} d^2\sigma/d\Omega dE_x$ (shaded region) is compared with $d^2\sigma/d\Omega dE_x$ (solid line). The calculation of Coulomb excitation (red circles) is also shown. The bin width is 0.2 MeV. \label{fig:spin}}
\end{figure}

\subsection{B. Decomposition of the cross section into spin-flip and non-spin-flip components}
We now discuss the energy spectra shown in Fig.~\ref{fig:cross} in more details. 
In a previous experiment~\cite{Tamii}, the polarization transfer (PT) observables were measured for  $\rm^{12}C$(\textit{p,p}$^\prime$)  at the same beam energy and 0$ ^{\circ}$ in the GR spectrometer, in which the excitation strengths were decomposed into a spin-flip part  ($\Delta S=1$) and a non-spin-flip part ($\Delta S=0$). Figure~\ref{fig:spin}(a) shows the cross section $d^2\sigma/d\Omega dE_x$ (solid line), the same as that in Fig.~\ref{fig:cross}, and the spin-flip cross section $\Sigma \hspace{0.1cm} d^2\sigma/d\Omega dE_x$ (shaded region). The total spin transfer  $\Sigma$  is unity for spin-flip transitions  ($\Delta S=1$) and zero for non-spin-flip transitions  ($\Delta S=0$). We used the $\Sigma$ values measured in the previous experiment~\cite{Tamii}, whereas the cross sections $d^2\sigma/d\Omega dE_x$ are our measurements. In the spin-flip cross section, excited states at $E_x$ = 18.35, 19.4, 22-23, and 25 MeV were observed whereas the non-spin-flip cross section was dominated by  broad resonances at $E_x$ = 22-24 and 25-26 MeV.  
 
 \subsection{C. Comparison of spin-flip cross sections with charge exchange reaction}
We now compare our $\Sigma \hspace{0.1cm} d^2\sigma/d\Omega dE_x$ (Fig.~\ref{fig:spin}(a)  shaded region) with the $T=1$ charge-exchange $\rm^{12}C$(\textit{p,n})$\rm ^{12}N$ spin-flip cross section measured at $E_p=296$ MeV~\cite{dozono}. The latter (\textit{p,n}) cross section was multiplied by  a factor of 0.5 (the Clebsch-Gordan coefficients) in order to compare with (\textit{p,p}$^\prime$) cross section. Moreover, the excitation energy was shifted for the case of the $\rm^{12}C$(\textit{p,n})$\rm ^{12}N$ reaction by 15.1 MeV.

The $T=1$ charge-exchange $\rm^{12}C$(\textit{p,n})$\rm ^{12}N$ spin-flip cross section was also measured at $E_p=135$ MeV by Anderson \textit{et al.}~\cite{anderson} and both data agree within the given errors.
Both observed resonances at $E_x$ = 19.4 ($2^-$), 22-23 ($2^-$), and 25 ($1^-$) MeV. 
Our spin-flip cross sections (shaded region) agree with the $T=1$ charge-exchange spin-flip cross sections, except for a small disagreement in the region $E_x = $ 18-19.4 MeV. This obvious disagreement arises from the fact that our data also includes isoscalar resonance at $E_x=$18.35 MeV, which is not observed in the charge exchange reaction.
This comparison primarily indicates that the (\textit{p,p}$^\prime $) spin-flip cross sections are mostly dominated by the $T=1$
 component, and the contribution of $T=0$ is small.
Indeed, the authors of Ref.~\cite{franey, petrovich} performed the analysis of the effective interaction ($V$) based on the \textit{N-N}
t-matrix for the nucleon-nucleus scattering data over the
energy range between 100 and 800 MeV.
They found that the spin-isospin term ($V_{ \sigma \tau} $
($T = 1$)) in the effective interaction  is much stronger than  the spin
term ($V_{ \sigma } $ ($T = 0$)) and that it is independent of the beam
energy.   
\begin{figure}[b]
\centering
\includegraphics[width=11.7cm,bb=0 0 1205 650]{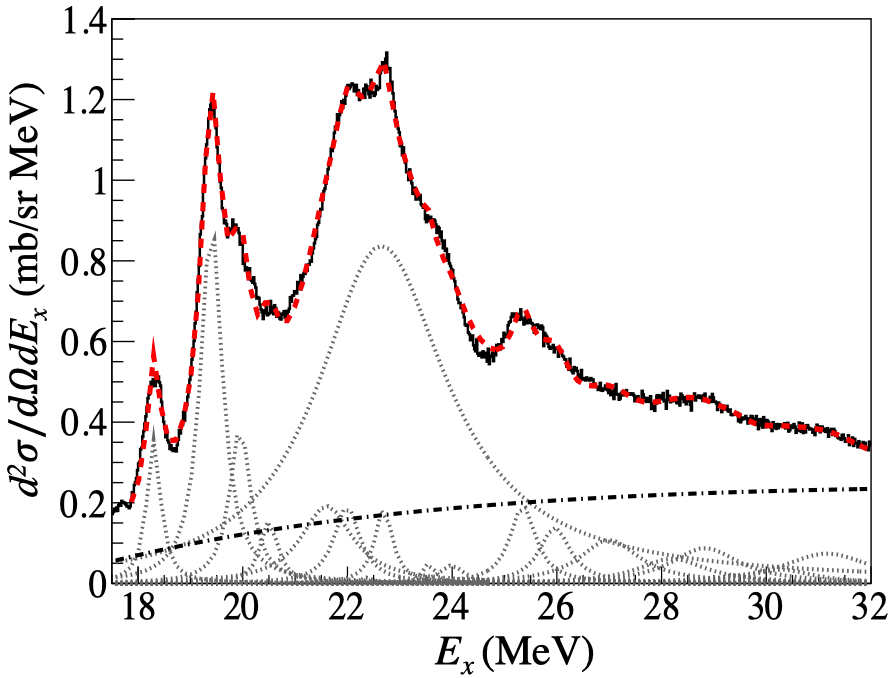}
\caption{Double differential cross section for the giant resonance region in $\rm^{12}C$ fitted with various resonances (dotted lines) \cite{kelley} and a quasifree continuum (dash-dotted line). The red dashed curve shows the overall fit obtained from the sum of all contributions.\label{fig:crossfit}}
\end{figure}

\begin{table}[h]
\begin{tabular}{>{\centering \arraybackslash}p{1.7cm}>{\centering \arraybackslash}p{1.3cm}>{\centering \arraybackslash}p{1.7cm}>{\centering \arraybackslash}p{2.0cm}} 
\hline
 \hline
$ E_m$ &  $J^{\pi};T$ & $\Gamma_m $ &$\sigma_m$ \\
 (MeV)              &                    &   (MeV)                     &(mb/sr MeV)\\
\hline
$18.35^{*1}$ 	 & $2^-$;0  	 &  0.35$\pm$0.05 &    0.35$\pm$0.03\\
19.40        	 & $2^-$;1  	 &  0.49$\pm$0.03 &   0.90$\pm$0.05\\
20.00          & $2^+$    	 &  0.38$\pm$0.10 &   0.39$\pm$0.04\\
$20.50^{*1}$ 	& $1^+$;0 	 & 0.30$\pm$0.05  &   0.15$\pm$0.03\\
21.60 			& $2^+$;0 	 & 1.20$\pm$0.15  &   0.18$\pm$0.02\\
21.99 			& $1^-$;1  	 & 0.61$\pm$0.11  &   0.19$\pm$0.06\\
22.37          & $1^-$;1  	 & 0.29$\pm$0.04  &   0.01$\pm$0.06\\
22.65 			& $1^-$;1 	 & 3.20$\pm$0.20  &   0.84$\pm$0.1\\
$22.68^{*2}$ 	& $1^-$;1 	 & 0.40$\pm$0.04  &   0.19$\pm$0.13\\
23.52	        & $1^-$;1 	 & 0.24$\pm$0.02 	& 0.06$\pm$0.06  \\
23.99 			& $1^-$;1 	 & 0.57$\pm$0.12  &   0.04$\pm$0.01\\
24.38 			& $2^+$;0	 & 0.67$\pm$0.06  &    0.00$\pm$0.00\\
24.41 			&      -       	 & 1.30$\pm$0.30  &   0.00$\pm$0.00\\
24.90 			&      -       	 & 0.90$\pm$0.20  &   0.00$\pm$0.00\\
25.30 			& $1^-$;1    & 0.51$\pm$0.10 &   0.19$\pm$0.04\\
25.40 			& $1^-$       & 2.00$\pm$0.20 &   0.00$\pm$0.00\\
25.96 			& $2^+$    	 & 0.70$\pm$0.20 &   0.14$\pm$0.02\\
27.00 			& $1^-$;1  	& 1.40$\pm$0.20  &   0.11$\pm$0.03\\
28.20 			& $1^-$;1  	& 1.60$\pm$0.20  &   0.06$\pm$0.01\\
28.83 			&      -        	& 1.54$\pm$0.09  &   0.09$\pm$0.01\\
29.40 			& $2^+$;1 	& 0.80$\pm$0.20  &    0.02$\pm$0.01\\
30.29 			& $2^-$;1 	& 1.54$\pm$0.09  &    0.04$\pm$0.01\\
31.16 			&      -       	 & 2.10$\pm$0.15 &    0.07$\pm$0.01\\
32.29 			&      -       	 & 1.32$\pm$0.23 &    0.01$\pm$0.01\\

\hline
quasifree continuum      &      -         &            -              & $\mu = 1.27\pm0.25$   \\
\hline
\hline
\end{tabular}
\caption{ Resonance energy ($ E_m $), resonance width ($ \Gamma_m$), spin-parity, and isospin obtained from Ref.~\cite{kelley}, and $\sigma_m$ obtained from fit.  *1) Spin-parity and isospin  were obtained from Ref.~\cite{johnson,Tamii}. *2) $E_m $ and $ \Gamma_m$  were obtained from Ref.~\cite{legge,bair}.}\label{tab:crossfit}
\end{table}  

\subsection{D. Comparison of non-spin-flip cross sections with $\rm^{12}C(\gamma,$\textit{total}) reaction}
  
 Figure \ref{fig:spin}(b) shows the cross section $d^2\sigma/d\Omega dE_x$ (solid line)  and  the non-spin-flip cross section $(1-\Sigma) \hspace{0.1cm} d^2\sigma/d\Omega dE_x$ (shaded region). 
It was suggested qualitatively by the $\rm^{16}O$(\textit{p,p}$^\prime$) experiment at the same beam energy (392 MeV) and 0$^{\circ}$~\cite{kawabata} that the non-spin-flip cross section is dominated by isovector giant dipole resonance ($J^{\pi}=1^-, T=1$) which is related to the Coulomb excitations.

We examined this feature more quantitatively by using the latest calculation of the Coulomb excitation~\cite{bertulani,peter} in the forward ($p,p^\prime$) reaction, which is expressed in terms of the total photo-nuclear absorption cross section~\cite{fuller}. The Coulomb excitation cross section was calculated at $1^\circ$  in Fig.~\ref{fig:spin}(b), since the average proton scattering angle was about  $1^\circ$. The calculation is shown in Fig.~\ref{fig:spin}(b) and agrees fairly well with the non-spin-flip data, except for the low energy region $E_x=$ 18-21 MeV and the high energy region $E_x>$30 MeV. In the low energy region our non-spin-flip  data also includes isoscalar resonance at  $E_x=$ 20.5 MeV which does not couple to the photo-absorption process and the data points are higher than the calculations. We also compared the calculation for Coulomb excitation with the non-spin-flip cross section for the $\rm^{58}Ni$(\textit{p,p}$^\prime$) reaction measured at $0^\circ$ in RCNP \cite{ishikawa} and found a good agreement within 10\%. Other small isoscalar contributions to the non-spin-flip cross section of $\rm^{12}C$ for $E_x$\textgreater 25 MeV were reported in a $\rm^{12}C$(\textit{d,d}$^\prime$) experiment~\cite{johnson} and a $\rm^{12}C(\alpha,\alpha^ \prime) $ experiment~\cite{itoh,kiss}.

\begin{table*}[!]

\begin{tabular}{>{\centering \arraybackslash}p{1.3cm}>{\centering \arraybackslash}p{1.1cm}>{\centering \arraybackslash}p{1.1cm}>{\centering \arraybackslash}p{1.1cm}>{\centering \arraybackslash}p{1.1cm}>{\centering \arraybackslash}p{1.1cm}>{\centering \arraybackslash}p{1.1cm}>{\centering \arraybackslash}p{1.1cm}>{\centering \arraybackslash}p{1.1cm}>{\centering \arraybackslash}p{1.1cm}>{\centering \arraybackslash}p{1.1cm}>{\centering \arraybackslash}p{1.1cm}>{\centering \arraybackslash}p{1.1cm}>{\centering \arraybackslash}p{1.1cm}} 
\hline
\hline
$  E_p $ & $ V $ &$ r_0 $ &$ a_0 $ &$W_v$ &$ r_0' $ &$a_0' $ & $V_{LS}$ & $r_{LS} $ &$ a_{LS} $ &$ W_{LS} $  &$ r_{LS}' $ & $a_{LS}'  $ &$r_{0C}$ \\
  (MeV) &  (MeV) &   (fm) &  (fm)&  (MeV) &   (fm) &  (fm)&  (MeV) &   (fm) &  (fm)&  (MeV) &   (fm) &  (fm)\\
\hline
398 & -2.51& 1.08 & 0.48 & 21.6 & 1.13 & 0.64 & 3.21 & 0.93 & 0.57 & -2.79 & 1.00 & 0.53 & 1.05\\
\hline
\hline
\end{tabular}
\caption{Optical model parameters used in DWBA calculations taken from Ref.~\cite{jones}.}\label{tab:optical}
\end{table*} 

\begin{table*}[!]
\centering
\begin{tabular}{>{\centering \arraybackslash}p{1.2cm}>{\centering \arraybackslash}p{1.1cm}>{\centering \arraybackslash}p{1.1cm}>{\centering \arraybackslash}p{1.2cm}>{\centering \arraybackslash}p{1.2cm}>{\centering \arraybackslash}p{1.2cm}>{\centering \arraybackslash}p{1.2cm}>{\centering \arraybackslash}p{1.2cm}>{\centering \arraybackslash}p{1.2cm}>{\centering \arraybackslash}p{1.2cm}>{\centering \arraybackslash}p{1.2cm}>{\centering \arraybackslash}p{1.2cm}} 
\hline
\hline
$ E_x$ & $  J^{\pi} ;T$ & $b$ & $ d_3p_1 $ &$ d_3p_3 $ &$d_5p_3  $ &$ s_1p_3 $ &$ d_5p_1 $&$ p_1p_1 $&$ p_1p_3 $&$ p_3p_1 $&$ p_3p_3 $\\
  (MeV)&&  (fm)\\
\hline
15.1$^{  (a)}$ & $ 1^+ ;1 $ &1.86 &-&-&-&-&-&-0.0581 &-0.6901 &-0.3394&-0.0764\\
15.1$^{  (b)}$ & $ 1^+ ;1 $ &1.86 &-&-&-&-&-&0.0829 &0.6701 &0.2904&0.0841\\
19.4$^{  (b)}$& $ 2^- ;1 $&1.64&-&-0.0926&0.5415&0.3043&-0.3047&-&-&-&-\\
22.8$^{  (b)}$&$ 1^- ;1 $ &1.64 &-0.1263&0.1472&-0.6874&-0.2108&-&-&-&-&-\\
\hline
\hline
\end{tabular}
\caption{Transition matrix elements used in DWBA calculations. The superscript (a) denotes transition matrix elements from Cohen and Kurath~\cite{jones}  and (b) denotes matrix elements obtained from SFO Hamiltonian~\cite{suzuki1,suzuki2}. The amplitude for the component $ l_il_j $ represents an excitation from the $ l_j $ hole state to the $ l_i $ particle state. The subscripts on the single-particle orbitals represent the quantity 2j. Here, the $ 2s_{1/2} $ orbital is designated as $ s_1 $.}\label{tab:transition}
\end{table*}
 \begin{figure*}[!]
\centering
\includegraphics[width=30cm,bb=0 0 2150 500]{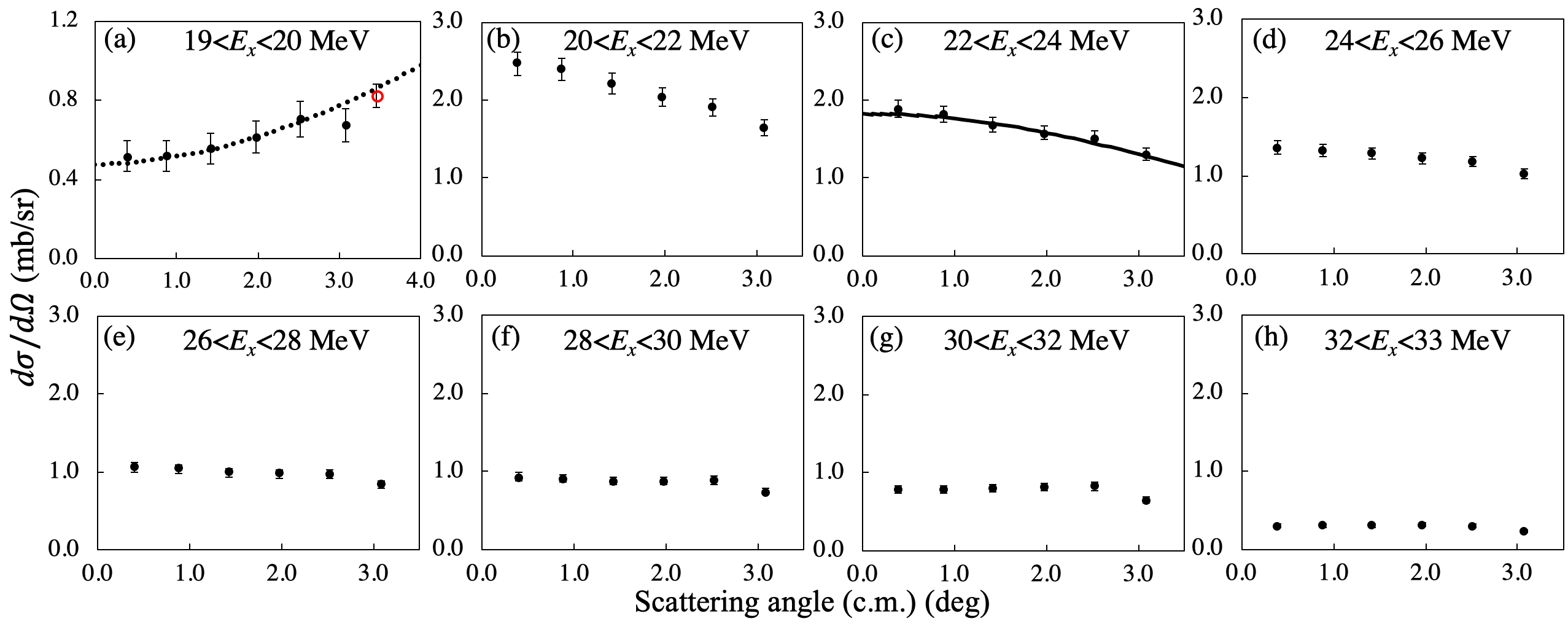}
\caption{ Differential cross section as a function of scattering angle at various excitation energies in the giant resonance region of   $\rm^{12}C$. Dotted and solid black lines show the result of DWBA calculations (see text). (a) A data point (red open circle) from another experiment \cite{jones} is also shown. \label{fig:dwba2}}
\end{figure*}

\subsection{E. Decomposition of different excited states}

It is clearly seen that the energy region $E_x$ = 16-32 MeV  consists of many overlapping resonances with different spin-parities and isospins. In order to unfold these  resonances, we fit the cross section with known resonances \cite{kelley} and a quasifree continuum.
The resonances were assumed to have Lorentzian distributions and the quasifree cross section was assumed to have a smooth functional form as described in Ref.~\cite{erell} (also shown in Fig.~\ref{fig:spin}(a)).  The overall fitting function was thus given as 
\begin{eqnarray}
f(E_x) = \sum_m \frac{\sigma_m}{1+(E_x^2-E_m^2)^2/E_x^2\Gamma_m^2} \nonumber \\ +\hspace{0.2cm} \mu  N  \frac{1-e^{[-(E_x-E_0)/T]}}{1+[(E_x-E_{QF})/W_L]^2},
\label{eqn:fraction}
\end{eqnarray} 
where $ E_m $ and $\Gamma_m$ are the peak energy and the resonance width, respectively, for the $m^{th}$ resonance. Their values were taken from  Ref.~\cite{kelley} and kept fixed during the fitting. The values of $N$ (0.2 mb/sr MeV), $ E_{QF}$ (27 MeV), $ W_L$ (55 MeV), $ E_0 $ (16 MeV), and $ T $ (6 MeV)  were determined from fitting to the $\rm^{12}C$(\textit{p,n})$\rm ^{12}N$ cross section \cite{dozono} and were kept fixed during this fit. The parameters $\sigma_m  $ (peak cross section)  and $\mu$  were determined to reproduce the data in the region of $ E_x =$ 18-32 MeV and are tabulated in Table \ref{tab:crossfit}. The fit is shown in  Fig.~\ref{fig:crossfit}. 

\subsection{F. Angular distribution in comparison with DWBA calculations}
We also present the differential cross section for the $\rm^{12}C$(\textit{p,p}$^\prime$) reaction as a function of scattering angle in various $E_x$ regions (Fig.~\ref{fig:dwba2}). Some of the angular distributions were compared with DWBA calculations.  

 The DWBA calculations were performed with the program DWBA07~\cite{DWBA07}. The single particle wave functions for the bound particles were of harmonic oscillator form. For the giant resonance region, the harmonic oscillator parameter $b$ = 1.64 fm was adopted~\cite{suzuki1,flanz}.  The distorted wave was derived by using an optical potential. The optical potential parameters were taken from  Ref.~\cite{jones}, as determined from 398-MeV proton scattering from $\rm^{12}C$, and are listed in Table~\ref{tab:optical}. The effective $NN$ interaction derived by Franey and Love~\cite{franey}  at $ E_p =425$ MeV was used. The transition densities were obtained from shell model calculations with SFO (Suzuki-Fujimoto-Otsuka) Hamiltonians~\cite{suzuki1,suzuki2} and are tabulated in Table~\ref{tab:transition}.

In Fig.~\ref{fig:spin}(a), it is clearly seen that the  energy region $E_x=$ 19-20 MeV is dominated by spin-flip cross section, and the data shown in  Fig.~\ref{fig:dwba2}(a) shows a clear angular dependence. The shape is well reproduced by the DWBA calculation results for the transitions to $E_x=$ 19.4 MeV ($ J^{\pi} =2^-, T=1$). For the energy region $E_x=$  22-24 MeV, which is dominated by Coulomb excitations, the calculation results for the transitions to $E_x=$ 22.8 MeV ($  J^{\pi} =1^-, T=1$) also reproduce the shape of angular distribution shown in Fig.~\ref{fig:dwba2}(c). For  $E_x$\textgreater 24 MeV, no clear angular dependence was observed.

We also tested DWBA for the cross section calculations of the 15.1-MeV state. The harmonic oscillator parameter was chosen~\cite{jones,comfort} to match the prominent maxima of longitudinal and transverse form factors  ($ F_L  (q) $ and $ F_T  (q) $) measured in a previous electron scattering experiment~\cite{flanz}. Two types of transition densities were used for the calculations of the 15.1-MeV state (Table~\ref{tab:transition}), the transition densities obtained from shell model calculations with SFO Hamiltonians~\cite{suzuki1,suzuki2} and 1-p shell transition densities from Cohen and Kurath~\cite{cohen,jones}. The comparison between calculations for these two different transition densities is shown in Fig.~\ref{fig:comp2}(a), along with the measured cross section. The dashed line represents the calculated cross section with transition densities from SFO Hamiltonians, and the solid line was obtained with Cohen and Kurath transition densities and was scaled by a factor of 1.15~\cite{jones}.

\section{IV. Analysis of emitted $\gamma$-rays}
\subsection{A. Definition and generation of response function $P (E_\gamma;E) $}

\begin{figure}[h!]
\centering
\includegraphics[width=8.5cm,scale=0.2]{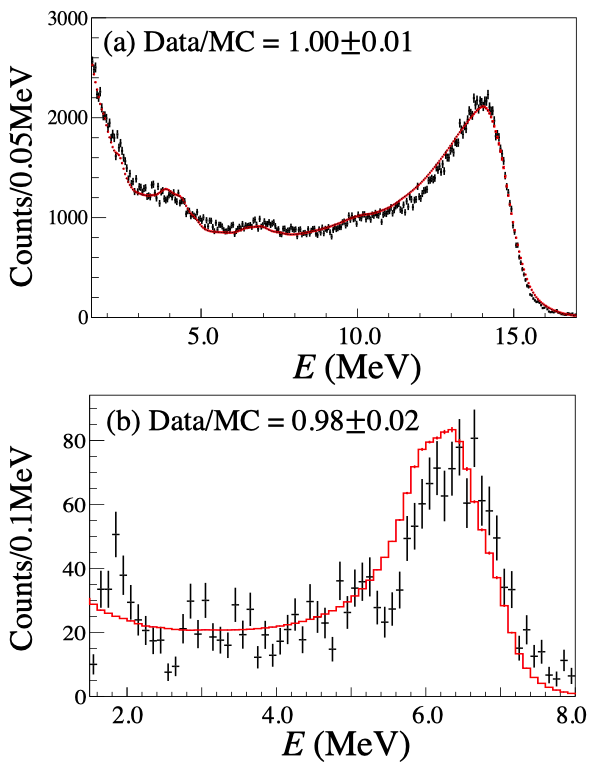}
\caption{Measured $\gamma$-ray spectrum (black data points) after  background subtraction and response function (red line) for the (a) 15.11-MeV state ($ J^{\pi}=1^+$) of $\rm ^{12}C$   (b) 6.9-MeV state ($ J^{\pi}=2^+$)  of $\rm^{16}O$ (see text for details). \label{fig:15MeV}}
\end{figure}

The response functions of the $\gamma$-ray detector were generated by geant4 Monte Carlo simulations (MC)~\cite{geant4}. The response function $P  ( E_\gamma;E) $ is defined as the probability for a $ \gamma $ ray of energy $ E_\gamma$  irradiated uniformly upon the target position to be measured as energy $E$ by the $ \gamma $-ray detector, and

\begin{eqnarray}
  \int_{E_{th}}^{E_{max}}P  ( E_{\gamma};E)dE  = \eta  ( E_{\gamma} ), 
\label{eqn:emitpd}
\end{eqnarray}
where $  \eta  ( E_\gamma) $ is the detection efficiency for a $ \gamma $ ray of energy $ E_\gamma$. For the present case, the threshold  ($  E_{th} $) for the $\gamma$-ray detectors was chosen to be 1.5 MeV. The detector geometry and the effect of the materials between the target and detector were taken into account during the detector simulation.  The accuracy of the  response functions was tested by comparison with the $\gamma$-ray spectra of 15.1 MeV and 6.9 MeV measured during the experiment.

\begin{figure*}[t]
\centering
\includegraphics[width=20cm,bb=0 0 1546 706]{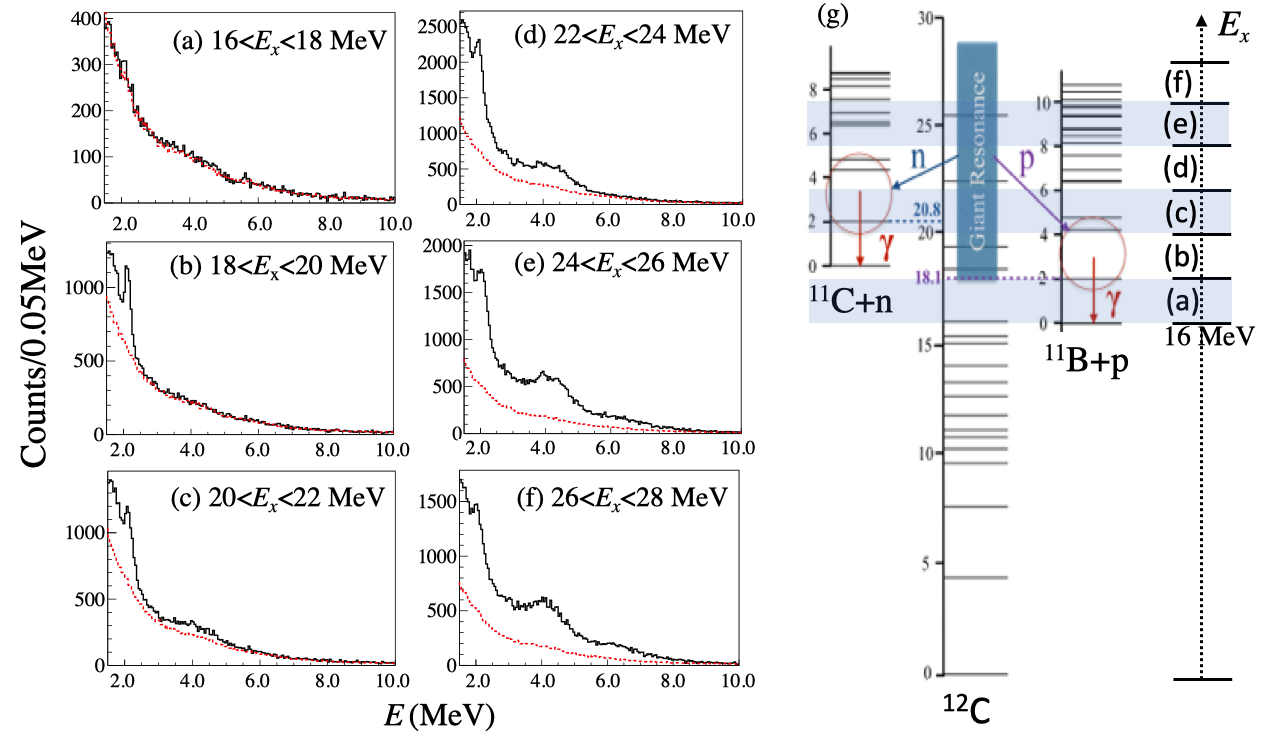}
\caption{ (a-f) $\gamma $-ray energy spectrum (black solid line) and background energy spectrum (red dashed line)  at various excitation energies in the giant resonance region of $ \rm^{12}C $. (g) The decay scheme of $\rm ^{12}C^*$.  \label{fig:gdecay}}
\end{figure*}

To generate the response function of a 15.1-MeV $\gamma$ ray, cascade $\gamma$ rays from the 15.1-MeV state~\cite{kelley}, 10.66, 7.45, 4.8, 4.4 and 2.4 MeV, were also taken into account, along with their respective branching ratios. The response function was then normalized by the 15.1-MeV excitation counts measured by the spectrometer in the energy range of $E_x=$ 14.9-15.4 MeV. Further, we determined the correction factor (0.88) for the response function to account for the dead time of the $\gamma$-ray detector by normalizing the data to reproduce the well-measured 15.1-MeV $\gamma$-ray emission probability ($\Gamma_\gamma / \Gamma = 0.96\pm0.04$). The response function for a 15.1-MeV $ \gamma $ ray is shown in Fig.~\ref{fig:15MeV}(a) (red line) along with the $\gamma$-ray energy spectrum measured  from the $\rm ^{12}C$ (15.1 MeV, $1^+$) (black points) after subtracting the background spectrum. The procedure for measuring the $\gamma$-ray spectrum and background subtraction was described in Section II(C) and shown in Fig.~\ref{fig:gammabg}. The photo peak and single- and double-escape peaks appear as one broad peak due to  the resolution of the $\gamma$-ray detector. This correction factor (0.88) was  used to scale the response function of all the other $ \gamma $ rays.

\subsection{B. Validation of response function $P (E_\gamma;E) $}

We show in  Fig.~\ref{fig:15MeV}(b) the $ \gamma $-ray spectrum (after background subtraction), as measured from  $E_x$($\rm^{16}O$) $=$ 6.9-7.3 MeV. Within this range, two states of $\rm^{16}O$, 6.9 MeV and 7.1 MeV were excited. These states decay to the ground state by emitting 6.9-MeV and 7.1-MeV $ \gamma $ rays, respectively, with 100\% emission probability.
The response functions were generated for 6.9 MeV and 7.1 MeV and weighted according to their contribution. A comparison with the response function  normalized by excitation counts in the same $E_x$ range is shown in Fig.~\ref{fig:15MeV}(b). When the value of data/MC for 15.1 MeV was normalized to 1.0  with the correction factor (0.88), the same factor yields data/MC = 0.98$\pm0.02$ for 6.9 MeV (including 7.1 MeV). The efficiency ($\eta(E_\gamma)$) was evalutated to be 2.3\% for $E_\gamma =$ 2.0 MeV and 5.9\% for $E_\gamma =$ 15.1 MeV.

 For the lower $ \gamma $-ray energy range, the consistency was checked  with a $\rm ^{60}Co$ source that emits two simultaneous $ \gamma $ rays with energies of 1.13 and 1.33 MeV. The response function generated for $\rm ^{60}Co$  reproduced the data within an uncertainty of 3\%.
The consistency between data and response function  within the systematic uncertainity of 5\% validates our measurement of $\gamma$-ray emission probability for the energy range from 1.1 to 15.1 MeV.

\begin{figure*}[t]
\centering
\includegraphics[width=19 cm,bb=0 0 1400 820]{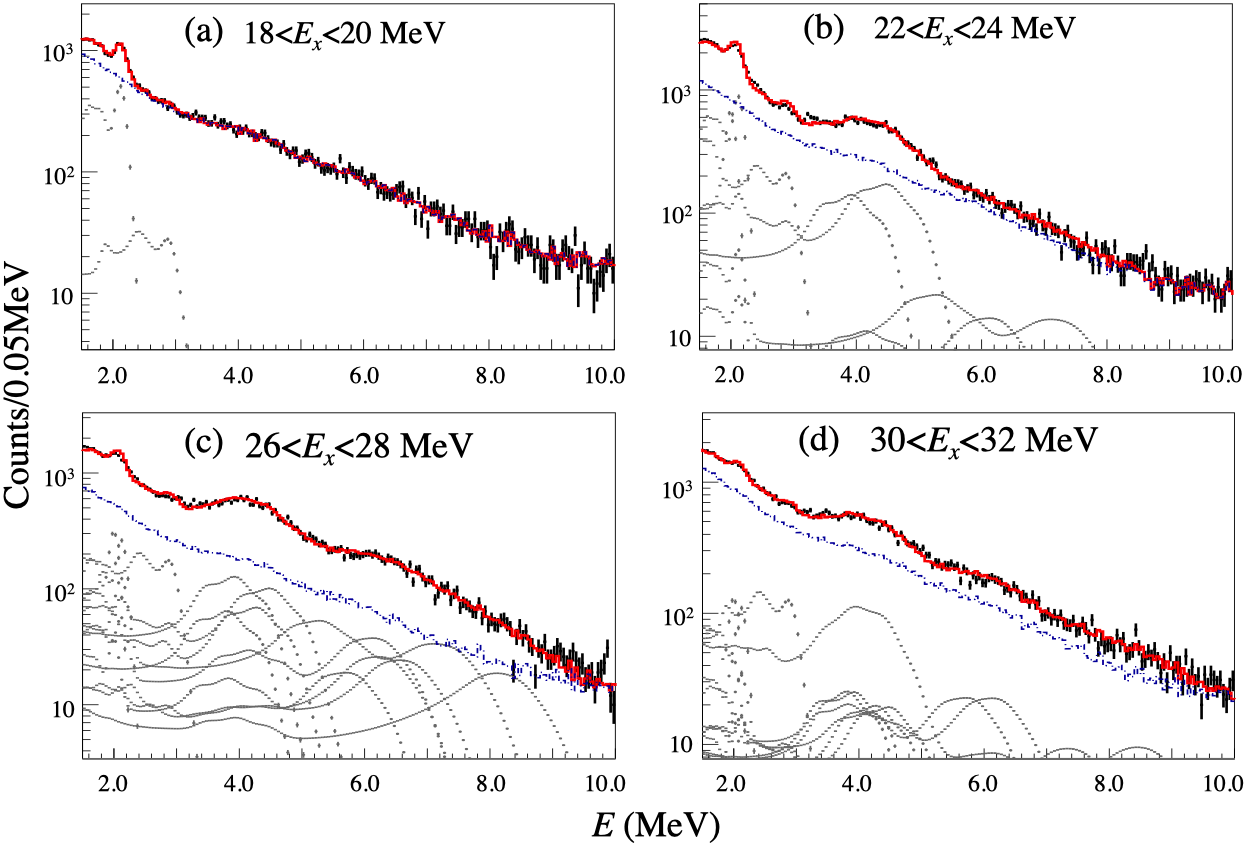}
\caption{The $\gamma$-ray spectrum (black data points), background spectrum (blue dash-dotted line), total fit (red solid line), and  $ \gamma $ rays from the excited states of the daughter nuclei  (grey dotted lines) are shown for various $E_x$ regions.\label{fig:fit}}
\end{figure*}

\section{V. $ \gamma $ rays from the giant resonances}

\subsection{A. $ \gamma $-ray energy spectra for each $E_x$ bin}
The $\gamma$-ray energy spectra from the giant resonances were measured for various $E_x$ values  with a 2-MeV energy step. Figure \ref{fig:gdecay} (left) shows the measured $ \gamma $-ray energy spectrum (black line) and background spectrum (red line). The decay scheme of excited $\rm ^{12}C$ is also shown. 

As  $E_x$ reaches  the proton separation energy  ($  S_p=16.0$ MeV),  the $\rm ^{12}C$ state decays hadronically to the ground state of $\rm ^{11}B$  by emitting a proton. No $\gamma$-ray emission is possible until  $E_x$ exceeds the threshold ($ S_p+2.1=18.1$ MeV)  for  proton decay to the first excited state of  $\rm ^{11}B$*(2.1 MeV). This feature was confirmed experimentally as no $\gamma$ rays were observed from the region $E_x=$16-18 MeV (shown in Fig.~\ref{fig:gdecay}(a)). The same feature can be seen in Fig.~\ref{fig:gdecay}(b) where we observed only a 2.1-MeV $\gamma$ ray, as the 2.1-MeV state of  $\rm ^{11}B$ is the only energetically accessible state at  $E_x=$18-20 MeV. As $E_x$ reaches 21 MeV, the $\rm ^{12}C$ state  can decay to the  $ 2^{nd}$ (4.4 MeV) and $3^{rd}$ (5.0 MeV) excited states  of $\rm ^{11}B$  or to the first excited state of  $\rm ^{11}C$*(2.0 MeV), after neutron emission ($ S_n$+2.0=20.7 MeV, $ S_n=$ 18.7 MeV). As a result, we observed nearly  doubled  $\gamma$-ray emission rate in Fig.~\ref{fig:gdecay}(c). 
With increasing  $E_x$, the larger $\gamma$-ray emission rate and  higher energy $\gamma$ rays were observed until the excitation energy reached 27.2 MeV, which is the separation energy of the daughter nuclei $\rm ^{11}B$  ($S_{p'}= 11.2$ MeV) and $\rm ^{11}C$  ($ S_{p'}= 8.7$ MeV). For $E_x$\textgreater 27.2 MeV,  the $\rm ^{12}C$ state can decay via  3-body decay to lighter nuclei. As far as hadronic decays are concerned, no $\gamma$ rays with $ E_{\gamma} $\textgreater 11 MeV were observed\footnote{The study of electromagnetic decay of giant resonances in $\rm ^{12}C$, emitting $\gamma$ rays of $E_\gamma$\textgreater 11 MeV, will be reported elsewhere.}.
These features agree qualitatively with the theoretical predictions of Langanke \textit{et al.}~\cite{Langanke,kolbe}, which states that the $\gamma$ rays from the giant resonances are emitted from  the excited states of the daughter nuclei after hadronic decay. We will further analyze the $\gamma$-ray emissions quantitatively.  

\begin{table}[ht]
\begin{tabular}{>{\centering \arraybackslash}p{1.85cm}>{\centering \arraybackslash}p{2.05cm}|>{\centering \arraybackslash}p{1.85cm}>{\centering \arraybackslash}p{2.05cm}} 
\hline
 \hline
Energy state & $\gamma$-ray energy & Energy state &$\gamma$-ray energy  \\
 ($\rm ^{11}B$)(MeV)              &   (MeV)(Prob.)                  & ($\rm ^{11}C$)    (MeV)                  &(MeV)(Prob.) \\
 \hline  \hline
 2.12  & $\gamma_0$2.12(1.0) & 2.00  & $\gamma_0$2.00(1.0)    \\
  \hline
4.44  & $\gamma_0$4.44(1.0) & 4.32  & $\gamma_0$4.32(1.0)    \\
 \hline
5.02  & $\gamma_0$5.02(0.85)& 4.80  & $\gamma_0$4.80(0.85) \\
      & $\gamma_{2.12}$2.89(0.15)&       & $\gamma_{2.00}$2.80(0.15) \\
    
       \hline
6.79  &$\gamma_0$6.79(0.68)    & 6.34  & $\gamma_0$6.34(0.67) \\
      &$\gamma_{2.12}$4.66(0.28)&         & $\gamma_{2.00}$4.33(0.33) \\
      
      &$\gamma_{5.02}$1.77(0.04)&       &             \\
   
       \hline
       7.28  &$\gamma_0$7.28(0.88)& 6.90  & $\gamma_0$6.90(0.92) \\
      &$\gamma_{4.44}$2.84(0.05)&         & $\gamma_{4.32}$2.58(0.04) \\
      &$\gamma_{5.02}$2.26(0.07)&       &$\gamma_{4.80}$2.10(0.04) \\
      
      \hline
7.97  &$\gamma_0$7.97(0.43)& 7.49  & $\gamma_0$7.49(0.36) \\
         &$\gamma_{2.12}$5.85(0.49)&       & $\gamma_{2.00}$5.49(0.64) \\
         &$\gamma_{7.28}$0.69(0.08)&       &             \\
     
      \hline
8.56  &$\gamma_0$8.56(0.56)& 8.10  & $\gamma_0$8.10(0.74) \\
         &$\gamma_{2.12}$6.43(0.30)&       & $\gamma_{2.00}$6.10(0.26) \\
         &$\gamma_{4.44}$4.11(0.05)&       &             \\
         &$\gamma_{5.02}$3.54(0.09)&       &              \\
     
      \hline
8.92  &$\gamma_0$8.92(0.95)& 8.42  & $\gamma_0$8.42(1.0)   \\
      &$\gamma_{4.44}$4.47(0.05)&       &               \\
  
      \hline
9.27  &$\gamma_0$9.27(0.18)& 9.20  & $\gamma_0$9.20(0.74)  \\
      &$\gamma_{4.44}$4.83(0.70)&       &$\gamma_{6.47}$2.72(0.20) \\
      &$\gamma_{6.74}$2.53(0.12)&       & $\gamma_{4.32}$4.88(0.13) \\
      
\hline
\hline
\end{tabular}
\caption{Energy states, $\gamma$-ray energies and emission probabilities of the daughter nuclei, where we follow the notation used in Table of Isotope \cite{TableIsotope}. Energy of the deexciting transition is preceded by $\gamma_{a}$ where $a$ is energy of the level populated by that transition. The emission probabilities of one energy state (given in parentheses) are normalized to 1.0 \cite{kelley}.}\label{tab:stategammarays}
\end{table}  

\begin{table*}[ht]
\begin{tabular}{ccccccccc}
\hline
\hline
 & & \multicolumn{7}{c}{$\rm ^{12}C$ Excitation Energy   $(E_x)$ (MeV)}\\
 \cline{3-9}
 &Energy state &18-20&20-22&22-24&24-26&26-28&28-30&30-32\\
 \cline{3-9}
 Decay Scheme&(MeV) ($J^\pi$)  &\multicolumn{7}{c}{$r_i$ [ \%]}\\
 \hline
\hline

 $\rm ^{11}B$+p & 2.12 ($1/2^-$)  &7.6(2)  &4.1(2) &9.2(2) &8.3(3) &5.9(3) &3.6(3)&2.8(4)  \\  

 ($S_p$=16.0 MeV)& 4.44 ($5/2^-$)  &- &1.0(2) &3.0(2) &5.9(3) &5.9(3) &2.4(3) &1.2(4)  \\  
                           
                           & 5.02 ($3/2^-$)   &- &1.2(2) &4.6(2) &5.7(3) &5.4(4) &2.9(5) &0.6(5)  \\

                           & 6.79 ($1/2^+$)       &- &- &0.6(1) &4.3(4) &3.2(5) &3.2(6) &2.3(3)  \\

          					&7.28 ($5/2^+$)    &- &- &- &0.8(4) &1.7(3) &0.5(3) &0.4(3)  \\

 							&7.97 ($3/2^+$)    &- &- &0.9(1) &2.9(5) &4.5(5) &- &-  \\

							&8.56 ($3/2^-$)    &- &- &- &1.9(3) &- &2.9(3) &1.0(2)  \\

							&8.92 ($5/2^-$)    &- &- &- &- &1.4(1) &- &-  \\

							&9.27 ($5/2^+$)    &- &- &- &- &- &2.8(7) &4.5(7)  \\  
          					  
          					\hline
$\rm ^{11}C$+n & 2.00 ($1/2^-$)   &- &2.4(1) &5.9(1) &6.5(2) &5.9(3) &3.6(3) &2.8(4)  \\ 

 ($S_n$=18.7 MeV)& 4.32 ($5/2^-$)   &- &- &- &1.0(1) &3.0(2) &2.4(3) &1.2(4)  \\  
                           
                           & 4.80 ($3/2^-$)      &- &- &- &3.2(2) &3.6(3) &2.2(4) &0.6(5)  \\

                           & 6.34 ($1/2^+$)      &- &- &- &- &1.6(2) &1.1(2) &2.3(4)  \\

                           & 6.90 ($5/2^+$)          &- &- &- &- &1.7(3) &0.5(3) &0.4(3)  \\

                           & 7.49 ($3/2^+$)        &- &- &- &- &- &- &-  \\

                           & 8.10 ($3/2^-$)          &- &- &- &- &- &1.5(1) &1.0(2)  \\

                           & 8.42 ($5/2^-$)         &- &- & -& -& -& -&-  \\ 
                            
                           & 9.20 ($5/2^+$)                                  &- &- &- &- &- &0.3(1) &0.5(1)  \\  
                                                 	
                            \hline	
         QF     &2.12 ($1/2^-$) &0.3(1)&0.9(2)&0.8(2)&1.4(3)&1.8(3)&2.2(3)&2.9(5)\\
                            &5.02 ($3/2^-$)  &-&0.3(1)&0.3(1)&1.0(2)&1.3(2)&1.7(2)&2.2(5)\\		
                            \hline	
                           &                        2.9               &0.8(2) &1.2(2) &4.2(2) &6.3(3) &7.5(4) &6.1(4) &6.0(4)  \\  
                          \hline
   \multicolumn{2}{c}{$R_\gamma (E_x)$ (\%)}             &8.4$\pm$0.5 &11.1$\pm$0.6 &28.6$\pm$1.6 &48.3$\pm$3.5 &53.3$\pm$3.9 &39.3$\pm$2.9 &33.3$\pm$2.5  \\            
      \hline \hline

\end{tabular}

\caption{The probability ($r_i$) obtained from the fit and the total $\gamma$-ray emission probability $R_\gamma (E_x)$ with systematic errors. Numbers in the parentheses represent the error in the least significant digit.} 
\label{tab:fitresults}
\end{table*}
 
 \subsection{B. Extraction of the $\gamma$-ray emission probability $R_\gamma (E_x)$ from the fit to the $\gamma$-ray spectra}
In order to obtain the $\gamma$-ray emission probability from the giant resonances of $\rm ^{12}C$, we fit the data with $\gamma$-ray response functions generated for the excited states of the daughter nuclei, which can be defined as

\begin{equation}
P_i(E)= b_0 \hspace{0.1cm} P(E_\gamma ^i;E) + \sum_{j=1} b_j \hspace{0.1cm} P(E_\gamma ^i - E_\gamma^j, E_\gamma^j; E ),
\end{equation}
where $P_i(E)$ is the response function for the $i^{th}$ state of the daughter nuclei at energy $E ^i$, $b_0$ is the probability for the $i^{th}$ state to decay directly to the ground state by emitting a $\gamma$ ray of energy $E_\gamma ^i$,  and  $b_j$  is the probability  for the $i^{th}$ state to decay to a lower energy  state ($E ^j$)  by emitting a $\gamma$ ray of energy $E_\gamma ^i - E_\gamma^j$ and then decay to the ground state by emitting  a $\gamma$ ray of energy $E_\gamma ^j $.  
For example, the first and the second excited states of $\rm ^{11}B$ decay directly to the ground state, emitting single $\gamma$ rays with energies of 2.12  and 4.4 MeV, respectively, with $b_0 = 1.0$. Hence, their response functions are given as \textit{P}(2.12 MeV;\textit{E}) and \textit{P}(4.4 MeV;\textit{E}). The third excited state of $\rm ^{11}B$ decays to the ground state by emitting a 5.02-MeV $\gamma$ ray with a probability of 0.85 ($b_0$) and  to the 2.12-MeV state by emitting a 2.9-MeV $\gamma$ ray ($5.02 -2.12$ MeV) with a probability of 0.15 ($b_1$) followed by further decay to the ground state by the emission of a  2.12-MeV $\gamma$ ray. The response function for this state is given as 0.85\textit{P}(5.0 MeV;\textit{E})+0.15\textit{P}(2.9, 2.12 MeV;\textit{E}). Similarly, the response function for all of the other excited states of the daughter nuclei ($\rm ^{11}B$ and $\rm ^{11}C$) were generated by using the $\gamma$ emission probabilities $(b_0$ and $b_j$) given in Ref.~\citep{TableIsotope} and are listed in 
Table~\ref{tab:stategammarays}. 
Once all of the response functions are generated, the efficiency ($\eta_i$) for the detection of $\gamma$ rays emitted from the $i^{th}$ state  of a daughter nucleus can be given as
\begin{eqnarray}
  \int_{E_{th}}^{E_{max}}P_i  ( E)dE  = \eta_i .
\label{eqn:emitpd}
\end{eqnarray}
The total $\gamma$-ray emission probability in each $E_x$ region of  $\rm ^{12}C$ can be written as
\begin{eqnarray}
  R_\gamma (E_x)  =\frac{\sigma _{p,p'\gamma}}{ \sigma _{p,p'}} = \frac{N_{\gamma}^0}{N_{E_x}},
\label{eqn:5}
\end{eqnarray}
where $ N_{E_x} $ is the total number of  excited states of  $\rm ^{12}C$ in that $E_x$ region  and $  N_\gamma ^0 $ is the total number of $ \gamma$ rays emitted from these states. The contribution from the individual excited states ($r_i$) of the daughter nuclei (after particle decay) to the total $\gamma$-ray emission probability can be given as
\begin{eqnarray}
  r_i    = \frac{N^0_i }{N_{E_x}} = \frac{ N  _i/\eta_i  }{N_{E_x}},
\label{eqn:7}
\end{eqnarray}
where $\ N^0_i   $ is the total number of $\gamma$ rays emitted from the $ i^{th}  $ state of the daughter nucleus from the target and $ N  _i $ is the number of events detected. The quantity $r_i$ can also be interpreted as the probability for  $\rm ^{12}C$ excited at $E_x$ to decay to the $i^{th}$ state of the daughter nuclei and emit a $\gamma$ ray. Furthermore, $r_i$ can be decomposed as  
\begin{eqnarray}
  r_i    = C_{GR}\hspace{0.1cm} \tilde{r}_i +C_{QF} \hspace{0.1cm} r_{QF}^i,
\label{eqn:7}
\end{eqnarray}
where $C_{GR}$ and $C_{QF}$ are the fractions of giant resonances (GR) and quasifree (QF) cross section in the total cross section obtained from Eq.~(\ref{eqn:fraction}), with
\begin{eqnarray}
 C_{GR}+C_{QF}= 1.0.
\label{eqn:7}
\end{eqnarray}
$\tilde{r}_i$ is the probability of giant resonance decaying to the $i^{th}$ excited state of the daughter nuclei and $ r_{QF}^i$ is the probability of the daughter nuclei to be in the $i^{th}$ excited state after quasifree knockout. The estimation of $\gamma$-ray emission probability  from quasifree process $ r_{QF}^i$ will be described in the next subsection C. The measured $\gamma$-ray spectrum ($  N_\gamma(E)$) in each $E_x$ region can be expressed as

\begin{eqnarray}
  N_\gamma (E) = N_{E_x}\sum_i r_i \hspace{0.1cm} P_i(E)  + {\alpha \hspace{0.1cm} N_{bg}(E)},                  
 \label{eqn:emitpd}
\end{eqnarray}
Alternatively, this can be written as
\begin{eqnarray}
   N_\gamma (E) = N_{E_x}\big{[} C_{GR}\sum_i \tilde{r}_i \hspace{0.1cm} P_i(E) \nonumber \\ + \  C_{QF} \sum_j r_{QF}^j \hspace{0.1cm} P_{j}(E)  \big{]}  + \  {\alpha \hspace{0.1cm} N_{bg}(E)},              
 \label{eqn:Nmeasured}
\end{eqnarray}
where  $ N_{bg}  (E)$ and $  N_{E_x} $ are  the background  spectrum and the number of excitation events, respectively. The quantities $  r_i $ and the background normalisation factor  ($ \alpha $) were set as free parameters in the fit.

\subsection{C. Estimation of $\gamma$-ray emission probability from quasifree processes}
The probability ($r_{QF}^j $)  after quasifree nucleon knockout can be obtained as follows.
A  proton knockout from the 1$p$ shell of $\rm^{12}C$  leads to  the $3/2 ^-$ ground state, the $1/2 ^-$ state at 2.1 MeV,  and the $3/2 ^-$ state at 5.02 MeV in $\rm^{11}B$. The spectroscopic factors for 1\textit{p} and 1\textit{s} knockout from $\rm^{12}C $  were experimentally determined from $\rm^{12}C$(\textit{e,e$^\prime$p})  data and are listed in Ref.~\cite{lapikas, steenhoven}.  Using 1\textit{p} spectroscopic factors,  the  probabilities for the daughter nucleus ($\rm^{11}B$ ) to be in  2.1-MeV  and 5.02-MeV states  were estimated to be ($r_{QF}^{2.12}=$) 4\%  and ($r_{QF}^{5.02}=$) 3\%, respectively. It should be noted that for $E_x$\textless 21 MeV, only the 2.1-MeV state is energetically accessible  with a probability of 4\%, but as $E_x$ exceeds 21 MeV, the 5.02-MeV state is also accessible. Similarly, a neutron knockout can also occur with equal probability and will lead to almost the same $\gamma$-ray response as that from a proton knockout. The only difference  is that the threshold for neutron knockout is greater than that for proton knockout by 2.7 MeV.

For $E_x$\textgreater 27.2 MeV, 1$s$ nucleon knockout can also occur. In this case, we used both 1$s$ spectroscopic factor and statistical model calculations (described in the next section) to estimate the contribution to the $\gamma$-ray emission probability. It was less than 1\% for $E_x$ = 27-32 MeV and was therefore ignored. 

Although 2.9-MeV $\gamma$ rays are expected from the decay of several states (5.02, 7.28 MeV, etc) and is included in their response functions, we found that an independent response function for  2.9 MeV must be added to Eq.~(\ref{eqn:emitpd}) to obtain a good fit. Furthermore, during the fit, 6.74-MeV ($7/2^-$) and 6.79-MeV ($1/2^+$) states of  $\rm ^{11}B$ and 6.48-MeV($7/2^-$) and 6.34-MeV($1/2^+$) states of $\rm ^{11}C$  were merged because these states lie  close to each other and were assumed to have the  same $ \gamma $-ray response function. Some of the fitted spectra are shown in Fig.~\ref{fig:fit}.

The total $ \gamma$-ray emission probability in different  ${E_x} $ regions can be given as  
\begin{eqnarray}
R_\gamma(E_x)  = \sum_i r_i   = C_{GR}\sum_i \tilde{r}_i +C_{QF} \sum_j r_{QF}^j,
\label{eqn:Rgamma}
\end{eqnarray}
This can  be equivalently written as
\begin{equation}
R_\gamma(E_x)  = \frac{(N_\gamma - N_{bg} )/\bar{\eta}}{N_{E_x}},
\label{eqn:total}
\end{equation}
where $N_\gamma$, $ N_{bg}$, and $  N_{E_x} $  are the number of  $ \gamma$-ray events, background events, and excitation events, respectively, and $\bar{\eta}$ is the weighted average efficiency in a particular $E_x$ region and $\bar{\eta}$ is given as 
\begin{eqnarray}
\bar{\eta} =  \frac{1}{\Sigma r_i} \sum_i r_i\hspace{0.1cm} \eta_i  \nonumber 
\end{eqnarray} 

\begin{eqnarray}
=\frac{1}{C_{GR}\sum_i \tilde{r}_i +C_{QF} \sum_j r_{QF}^j} \bigg{(} C_{GR}\sum_i \tilde{r}_i \hspace{0.1cm} \eta_i \nonumber \\ \ \ + C_{QF} \sum_j r_{QF}^j \hspace{0.1cm} \eta_j  \bigg{)}.  
\end{eqnarray}
The total $ \gamma$-ray emission probability  and the probability ($r_i$) obtained from the fit  are shown in Table~\ref{tab:fitresults} for all $E_x$ regions.

\begin{figure}[h]
\centering
\includegraphics[width=17cm,bb=0 0 1205 500]{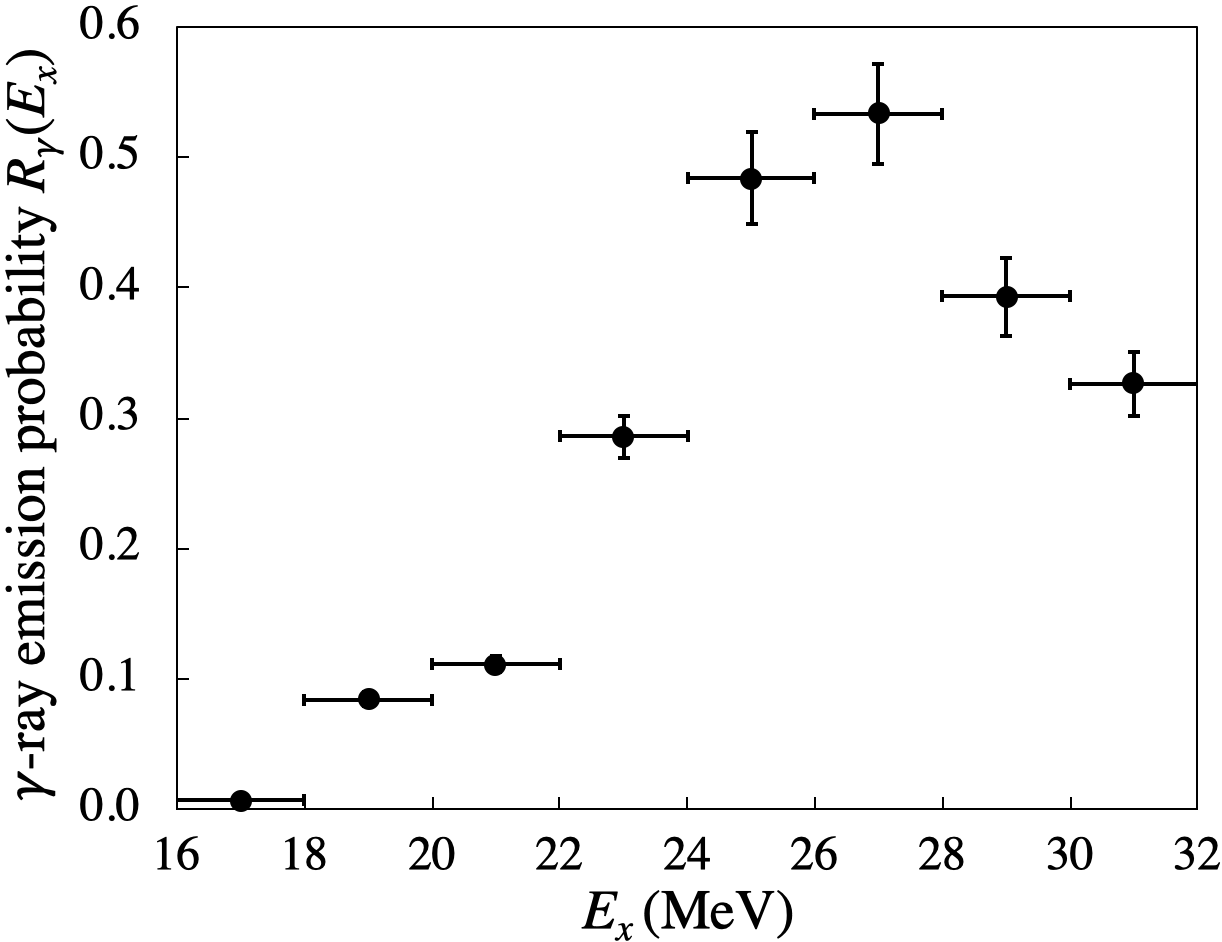}
\caption{Total $\gamma$-ray emission probability $ R_\gamma (E_x)$ as a function of $E_x$. The error bars include both statitstical and systematic  uncertainties.\label{fig:prgr12C}}
\end{figure}

\section{VI. Results of $\gamma$-ray emission probability $R_\gamma (E_x)$  and discussion}
\subsection{A.  $\gamma$-ray emission probability}

The $ \gamma $-ray emission probability  $ R_\gamma (E_x)$ as a function of excitation energy  ($E_x$) is shown in Fig.~\ref{fig:prgr12C} along with both statistical and systematic errors. 
The systematic uncertainties include the errors in the determination of excitation events (2-3\%), $ \gamma $-ray background subtraction (1-3\%), and  detection efficiency (5-7\%). The errors due to statistical uncertainty were 0.7-3\%.
The $ \gamma $-ray emission probability increases with the increasing excitation energy, starting from zero at $E_x = $ 16~MeV and reaches a maximum value of 53.3$\pm$0.4$\pm$3.9\% at $E_x=27$~MeV, where the first and second uncertainties are statistical and systematic, respectively. For $E_x$\textgreater 27 MeV, the emission probability gradually decreases with the increasing excitation energy. This feature is discussed later in detail. The most dominant contributions to the emission probability come from the 2.1 and 2.0-MeV states (first excited states of $\rm ^{11}B$ and $\rm ^{11}C$, respectively). For $E_x$\textgreater 26 MeV, the contributions of 8-9-MeV states of the daughter nuclei also become significant (Table \ref{tab:fitresults}).

\begin{figure*}[]
\includegraphics[width=30cm,bb=0 0 2180 600]{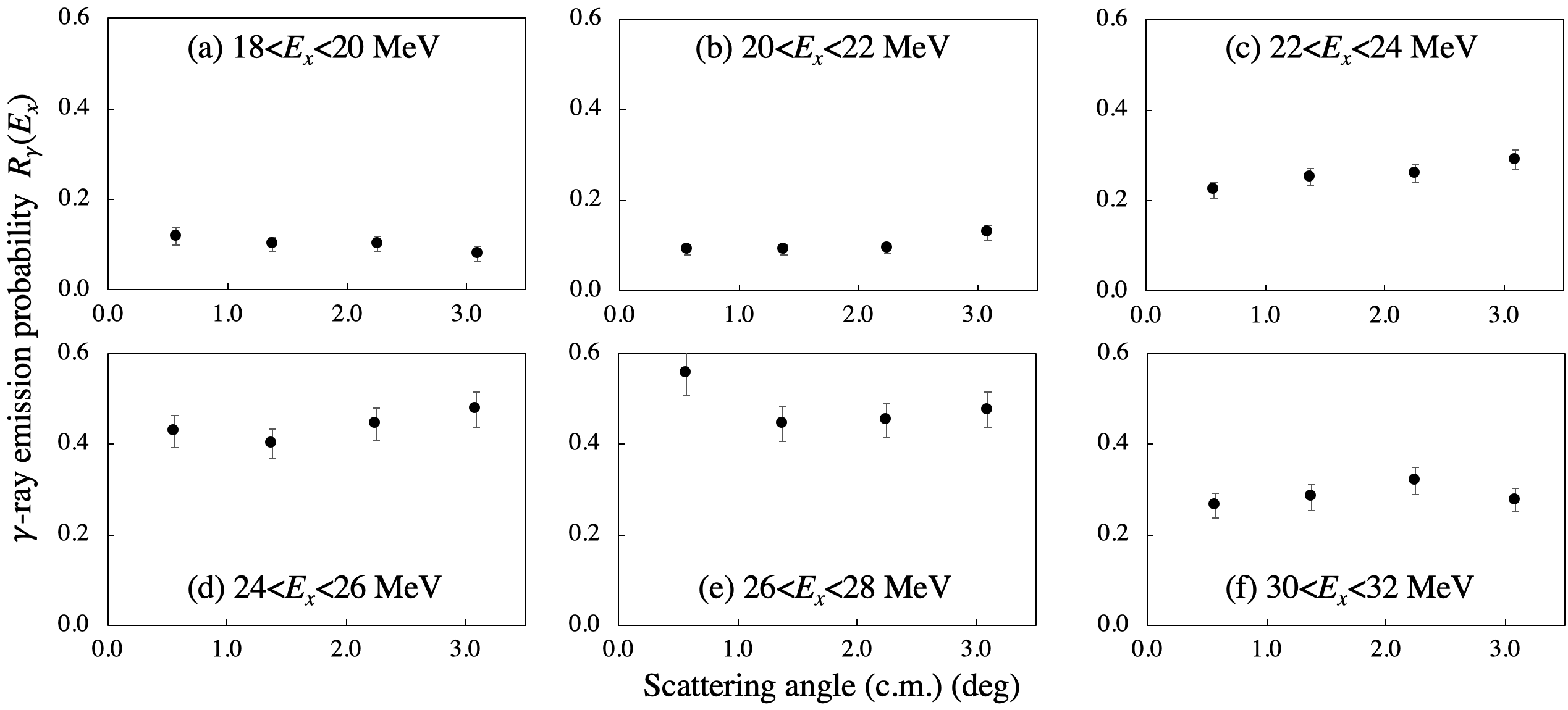}
\caption{ The $\gamma$-ray emission probability as a function of scattering angle at various excitation energies in the  giant resonance region of $\rm ^{12}C$.\label{fig:brangle}}
\end{figure*}

The $ \gamma $-ray emission probability was also measured as a function of scattering angle for different $E_x$ regions and no strong angular dependence was observed (Fig.~\ref{fig:brangle}).

\subsection{B. Comparison with decay model prediction}
A statistical model calculation based on the Hauser-Feshbach formalism \cite{hf, rauscher} was used to predict the $\gamma$-ray emission probability from the giant resonances of $\rm ^{12}C$ and is described as follows. 
The transmission coefficient from an excited nucleus $  (E_x)$ to the $i^{th}$ energy state of  a daughter nucleus A$ (E_A^i,J_A^i,\pi_A^i)$ by the emission of particle $a$ is given by the summation over all quantum mechanically allowed partial waves,
\begin{equation} \label{eq:trans}
T(E_x \rightarrow a+ (A,i)) =  \sum_{S=|J_A^i-s_a|}^{J_A^i+s_a} \sum_{L=|J_x-S|}^{J_x+S} T_L^a(\epsilon_a),
\end{equation}
where $T_L^a(\epsilon_a)$ is the individual transmission coefficient of the particle $a$ with kinetic energy  $\epsilon_a $ given by $E_x-E_A^i-$ separation energy, spin $s_a$, and orbital angular momentum  $L$.  The summation over $ L$ is restricted by the parity conservation rule $\pi_x=\pi_a \pi_A^i (-1)^L$. These individual  transmission coefficients were obtained by solving the $\rm Schr\ddot{o}dinger$ equation with the optical potential for the particle nucleus interaction \cite{cascade,murthy}. We employed global optical potential parameters given in  Ref.~\cite{ntrans, ptrans, atrans, dtrans} for the calculations.

The decay of an excited nucleus can proceed via different channels $a =$ \textit{ p, n, d, t }and $\alpha$. Then, the  probability for  an excited nucleus ($E_x$) to decay to the $i^{th}$ state of the daughter nuclei  can be given as
\begin{equation} \label{eq:br}
\tilde{c}_i  = \frac{\beta_a \hspace{0.1cm} T(E_x \rightarrow a+ (A,i)) }{\sum_{a,i} \beta_a \hspace{0.1cm} T(E_x \rightarrow a+ (A,i))},
\end{equation}
where $ \beta_a $ is the isospin Clebsch Gordan coefficient \cite{grimes,harakeh}. We used the spin-parity informations of Table~\ref{tab:crossfit}  for  the resonance states  in different $E_x$ regions and calculated the $\gamma$-ray spectrum $N_\gamma ^{calc} (E)$ as 
 \begin{eqnarray}
N_\gamma ^{calc} (E) = N_{E_x}\big{[} C_{GR}\sum_i \tilde{c}_i   \hspace{0.1cm} P_i(E) \nonumber \\ + \ C_{QF} \sum_j r_{QF}^j \hspace{0.1cm} P_{j}(E)  \big{]}  +  {\alpha \hspace{0.1cm} N_{bg}(E)}.\ \ \ \ \ \ \                   
 \label{eqn:Ncalc}
\end{eqnarray}
It should be noted that $\tilde{r}_i$ in Eq. (\ref{eqn:Nmeasured}) is replaced by $\tilde{c}_i$ in Eq. (\ref{eqn:Ncalc}).
Accordingly, the calculated $ \gamma$-ray emission probability $R_\gamma^{calc}(E_x)$ can be determined as  
\begin{equation}
R_\gamma^{calc}(E_x)  = C_{GR}\sum_i \tilde{c}_i + C_{QF}\sum_j r_{QF}^j.
\label{eqn:total}
\end{equation}
\begin{figure}[h!]
\centering
\includegraphics[width=17cm,bb=0 0 1205 501]{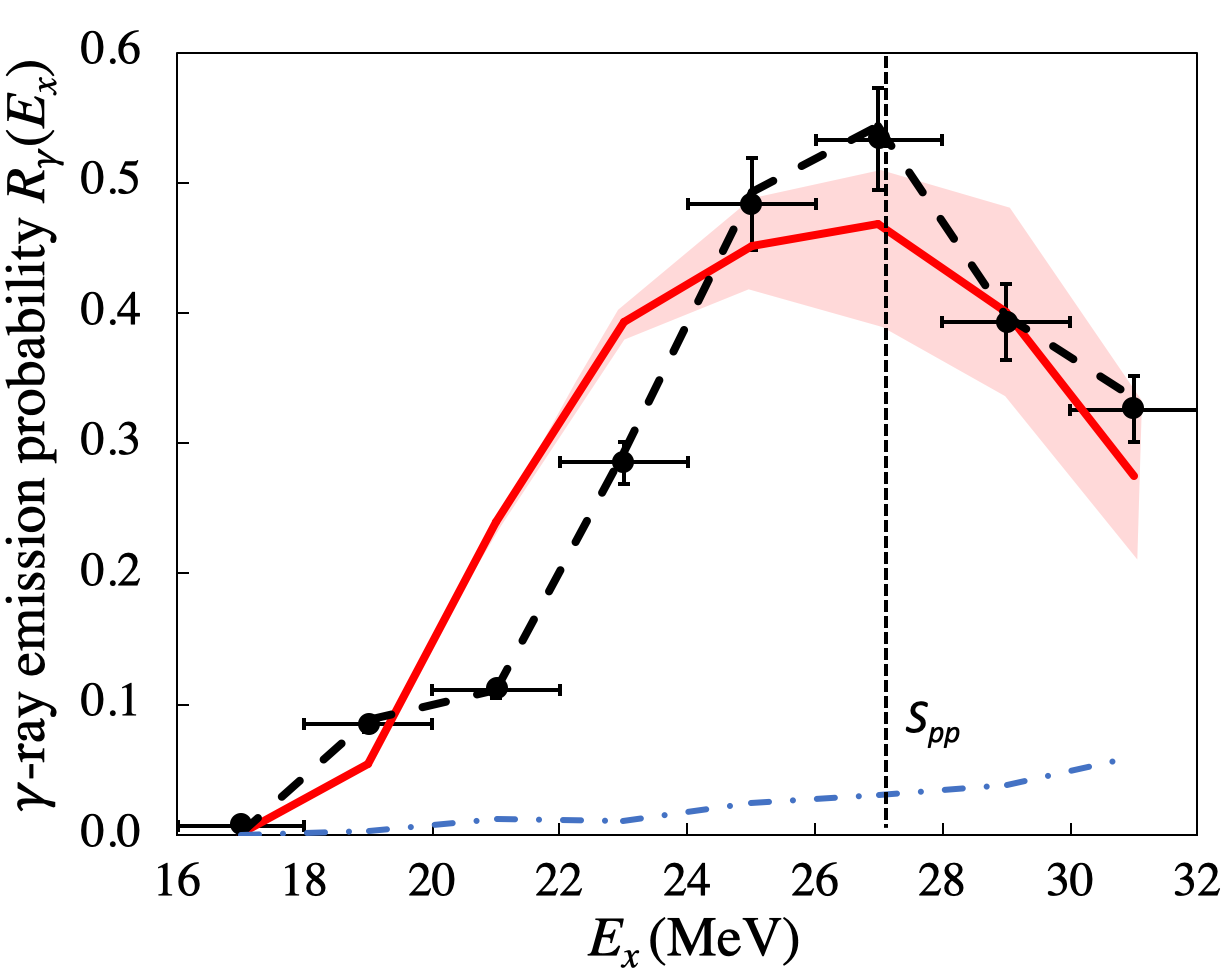}
\caption{ Comparison between the measured  $\gamma$-ray emission probability (data points) and the statistical model prediction (red solid line). The black dashed line shows the $\gamma$-ray emission probability obtained from the fit (Eq.~(\ref{eqn:Rgamma})). The red band shows the uncertainty in calculation due to the error in  $C_{QF}$. The $\gamma$-ray emission probability from quasifree process (blue dash-dotted line) is also shown. The quantity $S_{pp}$ represents two proton emission threshold (27.2 MeV) for $\rm^{12}C$.}
 \label{fig:cascade}
\end{figure} 
This probability is also shown in  Fig.~\ref{fig:cascade} as a red (solid) line. The $ \gamma $-ray emission probability from the quasifree process is also shown (blue dash-dotted line). 


The main contribution to the total $\gamma$-ray emission probability ($R_\gamma^{calc} (E_x)$) comes from the decay of giant resonances. For $E_x=$ 16-27 MeV, $R_\gamma^{calc} (E_x)$ increases because $C_{GR}$ dominates in this energy region and the number of accessible states of the daughter nuclei also increases. For $E_x$\textgreater 27 MeV, $C_{GR}$ begins to decrease and so does the $\gamma$-ray emission probability, while the contribution of $C_{QF}$ becomes nearly equal to $C_{GR}$. The red band in Fig.~\ref{fig:cascade} shows the uncertainty in the calculation due to the uncertainty in $C_{QF}$ ($\mu = 1.27 \pm0.25)$.

The statistical model calculations predicted a higher decay probability to the excited states by 30-40\% as compared to the measured values in the energy region $E_x=$ 20-24 MeV. The same feature was observed, when we compared calculations with the measurement of $\rm^{12}C$($\gamma,total)$ and $\rm^{12}C$($\gamma,n_0)$ cross sections \cite{fuller}.

For $E_x$\textgreater 27.2 MeV, the 3-body decay threshold is reached, and the decay involving two-nucleon emission ($p+p+\rm^{10}Be$) also starts.  Although the decay via 3-body process was significant ($\approx$ 6\%), it gave negligible contribution (\textless 1\%) to the $\gamma$-ray emission probability.

\section{VII. Conclusion}

We measured the double differential cross section ($d^2\sigma /dE_x d\Omega$) for the $\rm^{12}C$(\textit{p,p}$^\prime$) inelastic reaction at 392 MeV and 0$ ^{\circ}$ for the energy range $E_x$ = 7-32 MeV. Furthermore, the cross section was decomposed into spin-flip ($\Delta S=1$) and non-spin-flip ($\Delta S=0$) components by using  polarization transfer (PT) observables measured previously at the same beam energy \citep{Tamii}. The spin-flip cross section was observed to be dominated by isovector resonances and the non-spin-flip cross section was dominated by $1^-$ resonances and agreed well with recent calculations of Coulomb excitations \citep{bertulani}.

 For the measurements of $\gamma$ rays from the giant resonances, the absolute values of the $ \gamma $-ray emission probability $R_\gamma(E_x)$  and the response functions were verified by using in-situ $\gamma$ rays (15.1 and 6.9 MeV) with an  accuracy of $\pm$5\% during the experiment.  This calibration procedure  made it possible  to measure $R_\gamma(E_x)$ reliably as a function of the excitation energy of $\rm ^{12}C$ in the energy range $E_x$ = 16-32 MeV.
We found  that the measured value of $R_\gamma(E_x)$ starts from zero at $E_x=$~16~MeV (the threshold of $p+\rm^{11}B$ decay) and increases to  53.3$\pm$0.4$\pm$3.9\% at $E_x=27$ MeV and begins to decrease with further increase in $E_x$.

We compared the measurements of $\gamma$-ray emission probability with a statistical model calculation to understand our measured values. For $E_x = $ 16-27 MeV, the $\gamma$-ray emission probability increases with excitation energy because this energy region is dominated by giant resonances  and the number of accessible states of the daughter nuclei also increases. For $E_x$\textgreater 27 MeV, the dominance of giant resonances ceases and we observe the corresponding decrease in the $\gamma$-ray emission probability. In this energy region, the contribution from quasifree process to the total cross section becomes nearly equal to that of giant resonances, but still its total contribution to the $\gamma$-ray emission
probability is at most 5\% as shown in Fig. \ref{fig:cascade} (blue line). We also found that the contribution of 3-body decay process to the $\gamma$-ray emission probabililty was negligible. Quantitatively,
we observed a 30-40\% lower $ \gamma $-ray emission probability in the energy region $E_x=$ 20-24 MeV than that predicted by the statistical model calculation.

 The $ \gamma $-ray emission probability was also measured as a function of scattering angle, but no strong angular dependence was observed. 
 
The present results are very important for understanding the $\gamma$-ray emission probability of the giant resonances of a typical light nucleus ($\rm^{12}C$) and for the neutrino detection in liquid scintillator detectors through neutral-current interactions. A similar analysis of the  $\rm^{16}O$(\textit{p,p}$^\prime$) reaction is ongoing and will be presented elsewhere. An experiment with a Germanium detector such as that of the CAGRA spectrometer at RCNP \cite{sullivan} will significantly improve the current understanding of the $ \gamma $-ray emission and decay of giant resonances by separating $ \gamma $ rays emitted from the daughter nuclei after proton and neutron decays.

\section{Acknowledgements}
We  gratefully acknowledge the outstanding efforts of the RCNP cyclotron staff for providing a clean and stable beam for our experiment. We would like to thank Prof. H. Toki, the former Director of RCNP, for encouraging us at the early stage of this experiment. We also thank Profs. M.N. Harakeh, Y. Suda, H. Chiba, and H. Sagawa for valuable discussions.
This work was supported by JSPS Grant-in-Aid for Scientific Research on Innovative Areas (Research in a proposed research area) No. 26104006.

\bibliographystyle{apsrev4-1}

\end{document}